\documentclass[12pt,notoc]{JHEP3}

\usepackage{amsmath,amssymb,euscript,array,cite}

\input{epsf}
\newcommand{\startappendix}{
\setcounter{section}{0}
\renewcommand{\thesection}{\Alph{section}}}
\newcommand{\Appendix}[1]{
\refstepcounter{section}
\begin{flushleft}
{\large\bf Appendix \thesection: #1}
\end{flushleft}}
\usepackage{epsfig}

\def\N{{\cal N}}

\def\Dbarslash{\,\,{\raise.15ex\hbox{/}\mkern-12mu {\bar D}}}
\def\Dslash{\,\,{\raise.15ex\hbox{/}\mkern-12mu D}}
\def\delslash{\,\,{\raise.15ex\hbox{/}\mkern-9mu \partial}}
\def\delbarslash{\,\,{\raise.15ex\hbox{/}\mkern-9mu {\bar\partial}}}

\newcommand{\EQ}[1]{\begin{equation} #1 \end{equation}}
\newcommand{\AL}[1]{\begin{subequations}\begin{align} #1
\end{align}\end{subequations}}
\newcommand{\SP}[1]{\begin{equation}\begin{split} #1
\end{split}\end{equation}}

\title{Large N gauge theories and topological cigars}
\author{Gaetano Bertoldi, Timothy J. Hollowood\\
Department of Physics,\\ University of Wales Swansea,\\
Swansea, SA2 8PP, UK.\\
E-mail: {\tt g.bertoldi, t.hollowood@swan.ac.uk} \\
} 
\preprint{SWAT/06/479}

\abstract{We analyze the conjectured duality between a class of double-scaling limits of
a one-matrix model and the topological twist of non-critical superstring backgrounds
that contain the $N=2$ Kazama-Suzuki $SL(2)/U(1)$ supercoset model. The untwisted backgrounds are holographically dual to double-scaled Little String Theories in four dimensions and to the large N double-scaling limit of certain supersymmetric gauge theories. The matrix model in question is the auxiliary Dijkgraaf-Vafa matrix model that encodes the F-terms of the above supersymmetric gauge theories. We evaluate matrix model loop correlators with the goal of extracting information on the spectrum of operators in the dual non-critical bosonic string. The twisted coset at level one, the topological cigar, is known to be equivalent to the $c=1$ non-critical string at self-dual radius and to the topological theory on a deformed conifold.
The spectrum and wavefunctions of the operators that can be deduced from the matrix model double-scaling limit are consistent with these expectations.}


\begin{document}

\section{Introduction}

Since the work of 't~Hooft \cite{TH1}, which suggests that the large $N$ limit
of four-dimensional $SU(N)$ gauge theory admits a weakly coupled string theory description, 
there has been considerable interest in trying to find concrete examples
of such a duality, with the hope of gaining insight and analytical control 
over non-perturbative phenomena like confinement and chiral symmetry breaking.

The AdS/CFT correspondence \cite{AdS} and its generalizations
are examples of such gauge/string dualities. 
On the string side, however, one is usually limited to the
supergravity approximation due to technical difficulties of dealing with 
Ramond-Ramond fluxes. Consequently, it is interesting to study examples 
where one has a better control over the dual string 
worldsheet theory and some of these hurdles can be overcome.

In \cite{BD,BHM}, the large $N$ limit of a wide class of four-dimensional
${\cal N}=1$ theories
in a partially confining phase was studied. 
It was found that the low-energy description
of the theory breaks down close to points in the parameter space
where some baryonic and/or mesonic states become massless.
Nevertheless, this can be cured by defining a large $N$ double-scaling
limit (DSL) where one approaches
the singularity by keeping the mass $M$ of these states fixed. 
This limit has several interesting features. For example, the conventional 't~Hooft limit 
leads to a free theory of colour singlet states where all interactions are suppressed by powers of $1/N$. 
In the present case, the large $N$ Hilbert space splits into two decoupled sectors, and one of them keeps 
residual interactions whose strength is inversely proportional to the mass $M$. 
This suggests that the dynamics of this sector has a dual string description 
where the string coupling is given by $1/N_{eff}\sim\sqrt{T}/M$ 
where $T$ is the tension of the confining string.  
For some of these models where supersymmetry 
is actually enhanced in the DSL
from ${\cal N}=1$ to ${\cal N}=2$ or even 
${\cal N}=4$, 
the dynamics of the interacting subsector can be described 
by a double-scaled Little String Theory or, via holography, 
by a non-critical superstring background 
with no Ramond-Ramond flux and an exactly solvable worldsheet theory
\cite{GK,GKP,Kutasov:1990ua,LSZLST,LST,holog}.

The above large $N$ duality proposals are based on an analysis of the $F$-terms
of the theory, which, following the results of Dijkgraaf and Vafa, 
is performed by means of an auxiliary matrix model \cite{Dijkgraaf:2002fc,DV3,DVPW}.
In \cite{BHM}, it was shown that these large $N$ double-scaling limits 
correspond to a double-scaling limit
of the auxiliary matrix model that is analogous to the double-scaling limits
considered in \cite{oldMMdsl} to study $c \leq  1$ systems coupled to 2d gravity.  
In particular, it was shown that the double-scaling limits are well-defined 
in higher genus as well and that the free energy of the matrix model scales as
\EQ{
F_g  \thicksim M^{2-2g} \ . 
}
Furthermore, it was argued in \cite{BertoldiNC}, on the basis of the Dijkgraaf-Vafa correspondence and previous studies 
\cite{ADKMV, OoguriVafa,MV,GV,GK,GKP}, that the $c \leq 1$ system 
defined by the matrix model DSL considered in \cite{BD} is dual to 
the topologically twisted version of the non-critical superstring backgrounds. 

In this paper, we pursue the study of the matrix models DSLs
introduced in \cite{BD}, focusing on 
those models where the large $N$ double-scaled theory has ${\cal N}=2$
supersymmetry. The goal is to verify the duality between the matrix model DSLs
and the topologically twisted non-critical superstring backgrounds.
This would be a further consistency check of the holographic duality between 
Little String Theories defined
in the proximity of Calabi-Yau singularities and non-critical
superstring backgrounds 
proposed in \cite{GK,GKP}.
To achieve this, in Section \ref{loops}, we evaluate loop correlators 
of the double-scaled matrix model in the planar limit.
The rationale here is that from 
these matrix model correlators one can extract genus zero correlation
functions of local operators in the dual $c \leq 1$ non-critical bosonic theory,
as was done in \cite{BanksDSS,MSS} in the case of minimal models
coupled to 2d gravity (see \cite{Matrixreviews} for a comprehensive
review). The tool we use is the algorithm developed in \cite{Eynard} to solve
the matrix model loop equations. This is a particularly useful technique
because the nature
of the singularities is such that the orthogonal polynomial technique
cannot be applied \cite{BHM}.

This is a preliminary step towards verifying 
that  the $c\leq1$ non-critical bosonic string 
indeed corresponds to the topologically twisted non-critical superstring background. 
The simplest DSL that can be defined is associated to a conifold 
singularity and the relative non-critical superstring background is the 
$\N=2$ 
supersymmetric coset $SL(2)_k/U(1)$ at level $k=1$ \cite{OoguriVafa,MV,GV}. 
In this case, the matrix model DSL is expected to capture the topological A-twist
of this background (the ``topological cigar''). The topological cigar
was shown to be equivalent to the $c=1$ non-critical bosonic theory at self-dual radius \cite{MV} 
(see \cite{NN,AshokMT,AshokT} for a recent analysis), 
which is equivalent to the topological B model on the deformed conifold \cite{GV}.
We will verify that the matrix model loop correlators are consistent with 
this picture.

The large $N$ double-scaling limits of \cite{BD} are defined in the neighbourhood 
of Argyres-Douglas-type singularities. In section \ref{N=2}, we will compare the 
double-scaled free energy of the matrix model at genus zero and the prepotential 
of the relevant ${\cal N}=2$ curve. In particular, we will
evaluate their third derivatives with respect to 
the glueball superfields and the ${\cal N}=2$
moduli and find precise agreement upon rescaling as one approaches the singularity.  
This is a further check of the duality in the planar limit. We will also suggest a precise relation 
between the higher genus terms of the double-scaled matrix model and the higher genus 
terms of the ${\cal N}=2$ Seiberg-Witten free energy in a neighbourhood of the
Argyres-Douglas singularity. The Seiberg-Witten partition function in the neighbourhood of 
an Argyres-Douglas singularity would then correspond to the matrix model partition 
function defined by the near-critical spectral curve.

\section{The matrix model double-scaling limit}\label{doublereview}

In this section, we will review the matrix model singularities and relative
double-scaling limits studied in \cite{BD,BHM}.
Consider an ${\cal N}=1$ $U(N)$ theory with a chiral adjoint field $\Phi$ and superpotential
$W(\Phi)$.
The classical vacua of the theory are determined by the stationary points
of $W(\Phi)$
\EQ{
W(\Phi)=N
\sum_{i=1}^{\ell+1}\frac{g_i}i \text{Tr}_N\,\Phi^i\  .
\label{bsup}
}
The overall factor $N$ ensures that the superpotential scales 
appropriately in the 't~Hooft limit.
For generic values of the couplings, we find $\ell$ stationary points at the zeroes
of 
\EQ{
W'(x) = N \varepsilon \prod_{i=1}^\ell (x-a_i) \  ,  \qquad   \varepsilon \equiv g_{\ell+1}  \  .
} 
The classical vacua correspond to configurations where each of the $N$ eigenvalues of
$\Phi$ takes one of the $\ell$ values, $\{ a_i \}$, for $i=1,\ldots,\ell$.
Thus vacua are related to partitions of $N$ where $N_i \geq 0$ eigenvalues
take the value $a_i$ with $N_1 + N_2 + \ldots N_\ell = N$.
Provided $N_i \geq 2$ for all $i$, the classical low-energy gauge group
in such a vacuum is 
\EQ{
\hat{G}_{cl} = \prod_{i=1}^\ell U(N_i)  \approx  \prod_{i=1}^\ell U(1)_i \times SU(N_i) \  .
}    
Strong-coupling dynamics will produce non-zero gluino condensates
in each non-abelian factor of $\hat{G}_{cl}$. If we define as $W_{\alpha i}$
the chiral field strength of the $SU(N_i)$ vector multiplet in the low-energy theory,
we can define a corresponding low-energy glueball superfield 
$S_i = -(1/32\pi^2) \langle Tr_{N_i} ( W_{\alpha i} W^{\alpha i} ) \rangle$ 
in each factor. Non-perturbative effects generate a superpotential
of the form \cite{CSW,CDSW,VY}
\EQ{
W_{eff}( S_1, \ldots, S_\ell ) =
\sum_{j=1}^\ell N_j ( S_j \log( \Lambda^3_j/S_j) + S_j ) + 2 \pi i \sum_{j=1}^\ell b_j S_j  \  ,
}  
where the $b_j$ are integers defined modulo $N_j$ that label inequivalent 
supersymmetric vacua. 

Dijkgraaf and Vafa argued that the exact superpotential of the theory  
can be determined  by considering a matrix model with potential $W(\hat\Phi)$
\cite{Dijkgraaf:2002fc,DV3} 
\EQ{
\int d \hat \Phi \, \exp \left(  - g_s^{-1} \ Tr \, W(\hat\Phi)   \right)
= \exp \sum_{g=0}^\infty F_g \ g_s^{2g-2}\ ,  
}
where $\hat\Phi$ is an $\hat{N} \times \hat{N}$ matrix in the limit $\hat{N} \to \infty$.
The integral has to be understood as a saddle-point expansion around a critical point 
where $\hat{N}_i$ of the eigenvalues sit in the critical point $a_i$. 
Note that $\hat{N}$ is not related to the $N$ from the field theory.
The glueball superfields are identified with the quantities 
\EQ{
S_i = g_s \hat{N}_i \ , \qquad S = \sum_{i=1}^\ell S_i = g_s \hat{N}   
\label{defs}}
in the matrix model and the exact glueball superpotential is 
\EQ{
W_{gb}( S_1, \ldots, S_\ell ) =
\sum_{j=1}^\ell N_j  \frac{\partial F_0}{\partial S_j} + 2 \pi i
\sum_{j=1}^\ell b_j S_j\ ,  
\label{gbsp}} 
where $F_0$ is the genus zero free energy of the matrix model in the planar limit.
   
The central object in matrix model theory is the resolvent 
\EQ{
\omega(x)=\frac1{\hat N}\text{Tr}\,\Big(\frac1{x-\hat\Phi}\Big)\ .
}
At leading order in the $1/\hat N$ expansion, $\omega(x)$ is valued on
the spectral curve $\Sigma$, in this case a hyper-elliptic Riemann
surface defined by the algebraic relation
\EQ{
y^2=\frac1{(N\varepsilon)^2}\big(W'(x)^2+f_{\ell-1}(x)\big)\ .
\label{mmc}
}
The numerical prefactor is chosen for convenience. In terms of this curve
\EQ{
\omega(x)=W'(x)-N\varepsilon y(x)\ .
}
In \eqref{mmc}, $f_{\ell-1}(x)$ is a polynomial of order $\ell-1$
whose $\ell$ coefficients are moduli that are determined by the
$S_i$. In general, the spectral curve can be viewed as a double-cover
of the complex plane connected by $\ell$ cuts. For the saddle-point of
interest only $s$ of the cuts may be opened and so only $s$ of
the moduli $f_{\ell-1}(x)$ can vary. Consequently $y(x)$ has $2s$
branch points and $\ell-s$ zeros:\footnote{Occasionally, for clarity, 
we indicate the order of a polynomial by a subscript.}
\EQ{
\Sigma:\qquad y^2=Z_m(x)^2\sigma_{2s}(x)
\label{mc}
}
where $\ell=m+s$ and 
\EQ{
Z_m(x)=\prod_{j=1}^m(x-z_j)\ ,\qquad
\sigma_{2s}(x)=\prod_{j=1}^{2s}(x-\sigma_j)\ .
}
The remaining moduli are related to the
$s$ parameters $\{S_i\}$ by \eqref{defs}
\EQ{
S_i=g_s\hat N_i=N\varepsilon\oint_{A_i}y\,dx\ ,
}
where the cycle $A_i$ encircles the cut which opens out around the
critical point $a_i$ of $W(x)$.

Experience with the ``old'' matrix model teaches us that double-scaling
limits can exist when the parameters in the potential are varied in
such a way that combinations of branch and double points come
together. In the neighbourhood of such a critical point,\footnote{We
have chosen for convenience to take all the double zeros $\{z_j\}$ 
into the critical region.} 
\EQ{
y^2\longrightarrow C Z_m(x)^2B_n(x)\ ,
\label{redc}
}
where $z_j,b_i\to x_0$, which we can take, without loss of generality, 
to be $x_0=0$. The double-scaling limit involves first taking $a\to0$
\EQ{ 
x=a\tilde x\ ,\qquad z_i=a\tilde z_i\ ,\qquad b_j=a\tilde b_j
\label{dsl}
}
while keeping tilded quantities fixed.
In the limit, we can define the near-critical curve
$\Sigma_-$:\footnote{For polynomials, we use the notation
  $\tilde f(\tilde x)=\prod_i(\tilde
  x-\tilde f_i)$, where $f(x)=\prod_i(x-f_i)$, 
$x=a\tilde x$ and $f_i=a\tilde f_i$.}
\EQ{
\Sigma_-:\qquad y_-^2=\tilde Z_m(\tilde x)^2\tilde B_n(\tilde
x)\ .
\label{ncc}
}
It was shown in \cite{BHM}, generalizing a result of \cite{BD}, that in the limit $a\to0$, in
its sense as a complex manifold, the curve $\Sigma$ factorizes as
$\Sigma_-\cup\Sigma_+$. The complement to the near-critical
curve is of the form
\EQ{
\Sigma_+:\qquad y_+^2=x^{2m+n}F_{2s-n}(x)\ .
}
where $F_{2s-n}(x)$ is regular when $a=0$.

\subsection{Engineering the double-scaling limit on-shell}

It is important to stress that the above 
singularities are obtained on shell \cite{BD,BHM}. In the context
of supersymmetric gauge theories, the moduli $\{S_i\}$ are fixed by
extremizing the glueball superpotential \eqref{gbsp}. It is not, {\it a priori\/},
clear whether a double-scaling limit can be reached
whilst simultaneously being on-shell with respect to the glueball
superpotential. We will now review the analysis of \cite{BD,BHM} and show that
suitable choices of the coupling constants $\{g_i\}$ do indeed allow
for a double-scaling limit on-shell with respect to the glueball
superpotential. In general, the
potentials required are non-minimal. However, this is irrelevant for
extracting the universal behaviour that only depends on the
near-critical curve \eqref{redc}.

So the problem before us is to show that the critical point can be
reached simultaneously with being at a critical point of the
glueball superpotential. It is rather difficult to find
the critical points of the latter directly. 
Fortunately another more tractable method consists of comparing
the matrix model spectral curve \eqref{mmc}, the ``$\N=1$ curve'', with
the Seiberg-Witten curve of the underlying $\N=2$ theory that results when the
potential vanishes. The latter has the form
\EQ{
y_\text{SW}^2=P_N(x)^2-4\Lambda^{2N}\ ,
}
where $P_N(x)=\prod_{i=1}^N(x-\phi_i)$. Here, $\{\phi_i\}$ are a set of
coordinates on the Coulomb branch of the $\N=2$ theory 
and $\Lambda$ is the usual scale of strong-coupling effects in the
$\N=2$ theory.

When the $\N=2$ theory is deformed by addition of the superpotential
\eqref{bsup}, it can be shown that a vacuum exists when the Seiberg-Witten
curve and the $\N=1$ curve represent the same underlying 
Riemann surface \cite{CSW,deBoer:1997ap,CIV}. 
In concrete terms this means that, on-shell,
\SP{
&y_\text{SW}^2=P_N(x)^2-4\Lambda^{2N}=H_{N-s}(x)^2\sigma_{2s}(x)\\
&y^2=\frac1{(N\varepsilon)^2}\big(W^{\prime}(x)^2
+f_{s+m-1}(x)\big)=Z_m(x)^2\sigma_{2s}(x)\ ,
\label{con}
}
In these equations,  $H_{N-s}(x)$, $\sigma_{2s}(x)$, $Z_{m}(x)$
are polynomials of the indicated order, and we choose
(in order to remove some redundancies)
\EQ{
H_{N-s}(x)=x^{N-s}+\cdots\ ,\qquad
\sigma_{2s}(x)=x^{2s}+\cdots,\qquad
Z_m(x)=x^m+\cdots\ .
}
Both curves describe the same underlying Riemann surface,
namely the reduced curve of genus $s-1$ which is a hyper-elliptic
double-cover of the complex plane with $s$ cuts.
All-in-all there are $2(N+l)$ equations for the same number of unknowns in
$\{P,H,\sigma,Z,f\}$. There are many solutions to these equations
and we can make contact with the description of the vacua in Section 1
by taking the classical limit $\Lambda\to0$; whence
\EQ{
P_N(x)\to\prod_{i=1}^{\ell}(x-a_i)^{N_i}\ ,\qquad\sum_{i=1}^{\ell}N_i=N\ ,
}
so $N_i$ of the eigenvalues of the Higgs field classically lie at the
critical point $a_i$ of $W(x)$. Quantum effects then have the effect
of opening the points $a_i$ into cuts (if $N_i>0$). The number of
$N_i>0$, {\it i.e\/}~the number of cuts, is equal to $s=\ell-m$.

We now to turn to explicit solutions of \eqref{con}. The method we
shall adopt is to first find solutions for a $U(p)$ gauge theory and
then apply the ``multiplication by $N/p$ map'' \cite{CIV}, with $N/p$ integer. 
This will yield a solution for a $U(N)$ gauge group and will allow to take a large $N$
limit with $p$ fixed.

\subsection{No double points}

We now describe how to engineer 
the case where the near critical curve \eqref{redc} has 
no double points, so $m=0$. This is the situation considered in 
\cite{BD,Eguchi:2003wv,bert}. 
In this case, we first consider the consistency conditions
\eqref{con} for a $U(p=n)$ gauge theory with $W(x)$ of order
$\ell=n+1$. In this case, \eqref{con} are trivially satisfied with
\EQ{
W'(x)=N\varepsilon P_n(x)\ ,\qquad f_{n-1}(x)=-4N^2\varepsilon^2\Lambda^{2n}\ .
\label{ndp1}}
Notice that with our minimal choice of potential, the on-shell curve
actually implies that $S=0$ since the coefficient of $x^{n-1}$ in
$f_{n-1}(x)$ vanishes and so the resolvent falls faster than $1/x$ at
infinity. This, of course, is pathological from the
point-of-view of the old matrix model and may be remedied by using a
non-minimal potential with extra branch points or double points
outside the critical region. However, in the holomorphic context in
which we are working, having $S=0$ is perfectly acceptable and we
stick with it.  
The on-shell curve consists of an $n$-cut hyperelliptic curve and one
can verify, by taking the classical limit, that $N_i=1$, $i=1,\ldots,n$. The
double-scaling limit involves a situation where $n$ branch points, one
from each of the cuts, come together. This can be arranged by having
\EQ{
W'(x)=N\varepsilon\big( B_n(x)
+2\Lambda^n\big)\ ,\qquad B_n(x)=\prod_{j=1}^n(x-b_j)
\label{cioppino}}
and then taking the limit \eqref{dsl}. In this case, the near-critical
curve $\Sigma_-$ \eqref{ncc} is of the form
\EQ{
y_-^2=\tilde B_n(\tilde x)\ .
}
The important point is that we can tune to the critical region whilst
keeping the theory on-shell with respect to the glueball
superpotential by simply changing the parameters $\{b_j\}$ which
appear in the potential.

Now that we have found a suitable vacuum of a $U(n)$ theory, we now
lift this to a $U(N)$ theory with the multiplication by $N/n$ map
\cite{CIV}. 
Under this map, the $\N=1$ curve remains intact, including the 
potential $W(x)$ whilst the Seiberg-Witten curve of the $U(N)$ theory is 
\EQ{
y_{SW}^2=P_N(x)^2-4\Lambda^{2N}=\Lambda^{2(N-n)}{\cal
  U}_{\tfrac Nn-1}\Big(\frac{P_n(x)}{2\Lambda^n}\Big)^2\big(P_n(x)^2-
4\Lambda^{2n}\big)\ ,
\label{mnm}
}
where ${\cal U}_{\tfrac Nn-1}(x)$ is a Chebishev polynomial of the second
kind. The vacuum of the $U(N)$ theory has $N_i=N/n$, $i=1,\ldots,n$. 

Notice that in the near critical region the Seiberg-Witten curve 
is identical to $\Sigma_-$, up to a rescaling:
\EQ{
y_\text{SW}^2\longrightarrow\Big(\frac{2N}{n}\Big)^2
\Lambda^{2N-n}B_n(x)\ .
\label{nhy}
}
This is
simply a reflection of the observation of \cite{BD} that the decoupled
sector has enhanced $\N=2$ supersymmetry. Moreover, if $C$ is a cycle
which is vanishing as $a\to0$ then the integral of the Seiberg-Witten
differential around $C$, which gives the mass of a BPS state carrying
electric and magnetic charges in the theory, becomes
\EQ{
\oint_C\frac{xP'_N(x)\,dx}{y_\text{SW}}
\longrightarrow - \Lambda^{-n/2} Na^{n/2+1}\oint_C y_-\,d\tilde x\ .
}
Notice that in the double-scaling limit \eqref{limitI1} (with $m=0$)
the mass of the state is fixed. This state is a dibaryon that carries electric 
and magnetic charges of the IR gauge group. In the
double-scaling limit, therefore, a set of mutually non-local dibaryons
become very light.\footnote{For $n=2$ there is only a single light dibaryon.}
In fact, the Seiberg-Witten curve at the critical point, $a=0$, has the form
\EQ{
y_\text{SW}^2=4\Big(\frac{N}{n}\Big)^2
\Lambda^{2N-n}x^{n}\ ,
}
which describes a ${\bf Z}_n$ or $A_{n-1}$ Argyres-Douglas singularity 
\cite{AD,ARSW,EHIY}.

\subsection{With double points}

For the case with double points, we cannot simply take two of the branch points
$\{b_j\}$ in \eqref{cioppino} above to be the same. If we simply did
that then the zero of the Seiberg-Witten curve, by which we mean a 
zero of the polynomial $H_{N-s}$ in \eqref{con}, would also be a zero 
of the $\N=1$ curve as well. By the analysis of \cite{deBoer:1997ap},
this would imply that the condensate of the associated massless dibaryon
is vanishing and the dual $U(1)$ group unconfined. 
On the contrary, we need to arrange a situation where any zero of the Seiberg-Witten
curve is not simultaneously a zero of the $\N=1$ curve, so that  
the putative massless dibaryon is condensed and the dual $U(1)$ is confined.

A suitable $\N=1$ curve which reduces to \eqref{redc} in the near-critical
region is
\EQ{
y^2=Z_m(x)^2B_n(x)\big(B_n(x)H_{r}(x)^2+4\Lambda^{2r+n}\big)\ .
\label{noc}
}
In this case, we have $\ell=m+n+r$, $s=n+r$ and
\EQ{
W'(x)=N\varepsilon Z_m(x)B_n(x)H_{r}(x)\ ,\qquad
f_{\ell-1}(x)=4N^2\varepsilon^2\Lambda^{2r+n}Z_m(x)^2B_n(x)\ .
}
Notice that in order that $f_{\ell-1}(x)$ has order less than $\ell$ we
require $r>m$. The curve \eqref{noc} is actually on-shell with respect
to the Seiberg-Witten curve of a $U(2r+n)$ theory with
\EQ{
P_{2r+n}(x)=H_{r}(x)^2B_n(x)+2\Lambda^{2r+n}\ .
}
In the classical limit, we have two eigenvalues at each of the
zeros of $H_r(x)$ and one in each of the zeros of $B_n(x)$.
Once again we can employ the multiplication map \eqref{mnm} (with $n$
replaced by $2r+n$) to find the vacuum of the $U(N)$ theory we are after. 

Notice that the double points of the Seiberg-Witten curve $\{h_i\}$
are not generally zeros of the curve \eqref{noc}, which means that the
associated dyons are condensed.
The near-critical curve $\Sigma_-$ in this case is
\EQ{
y_-^2=\tilde Z_m(\tilde x)^2\tilde B_n(\tilde x) \ ,
}
while in the near-critical region the Seiberg-Witten curve becomes
\EQ{
y_\text{SW}^2\longrightarrow4\Big(\frac{N}{n+2r}\Big)^2
\Lambda^{2N-2r-n}H_r(0)^2B_n(x)\ .
\label{swnc}
}
where we assumed that the zeros of $H_r(x)$ lie outside the critical
region. In this case, the integral of the Seiberg-Witten differential
around a vanishing cycle diverges in the double-scaling limit:
\EQ{
\oint_C\frac{xP'_N(x)\,dx}{y_\text{SW}}\thicksim Na^{n/2+1}=\Delta
a^{-m}\to\infty\ .
}
So in contrast to the case with no double points, 
the dibaryon states are very heavy. In addition, the
dyon condensate associated to the zero $h_i$ of $H_r(x)$ is given by an 
exact formula \cite{deBoer:1997ap} 
\EQ{
\langle m_i\tilde m_i\rangle=N\varepsilon y(h_i)\thicksim N
\to\infty\ ,
}
where we have assumed that $h_i$ stays fixed as $a\to0$.
So in the double-scaling limit the value of the condensate and hence
the confinement scale in the dual $U(1)$, or string tension, 
occurs at a very high mass scale. 

Even though there are no light dibaryons as in the previous
example, there is still an interesting double-scaling
limit in the gauge theory due to the presence of other light mesonic states in the theory
with a mass $\sim\Delta$ \cite{BHM}.

Notice also that contrary to our choice above, if we scale $h_i\to0$ as $a\to0$ then
the tensions of the confining strings vanish and the 
theory is at an $\N=1$ superconformal fixed point in the infra-red
corresponding to one of the $\N=1$ Argyres-Douglas-type singularities
described in \cite{Eguchi:2003wv}. As the double points of 
the Seiberg-Witten curve $h_i$
move away from the origin the associated dyons condense and the superconformal
invariance is broken. 

\subsection{Higher genus}

In the $a\to0$ limit, it was shown in \cite{BHM} 
that the genus $g$ free energy gets a dominant contribution from $\Sigma_-$
of the form
\EQ{
F_g\thicksim \big(Na^{(m+n/2+1)}\big)^{2-2g}\ .
\label{dsp}
}
Note that in this equation $N$ is the one from the field theory and
not the matrix model $\hat N$. This motivates us to define the 
double-scaling limit \cite{BD,BHM}
\begin{equation}
a \rightarrow 0 \ , \qquad N \rightarrow \infty
\ , \qquad \Delta \equiv N a^{m+n/2+1} = \text{const} \ .
\label{limitI1}\end{equation}
Moreover, the most singular terms in $a$ in \eqref{dsp} depend only on the
near-critical curve \eqref{ncc} in a universal way.

It was observed in \cite{BertoldiNC} that Eq. \eqref{dsp} matches the expected behaviour of the 
topological B model free energy at the singularity \cite{BCOV}.
In fact, as can be seen from \eqref{redc} and \eqref{dsl}
\EQ{
\Delta \sim N \int y\,dx   \  .
\label{olalala2}}
More precisely, the double-scaling parameter is proportional
to the period of the one-form $y\,dx$ on one of the cycles that 
vanish at the singularity.
Moreover, this one-form corresponds to the reduction
of the holomorphic $3$-form $\Omega$ on the underlying Calabi-Yau geometry
\AL{
uv + y^2 &= W'(x)^2 + f(x) \\
\Omega &= \frac{dudvdx}{\sqrt{uv - W'(x)^2 - f(x)}} \ .
}
This comes from the fact that $3$-cycles in the Calabi-Yau correspond
to ${\boldsymbol S}^2$ fibered over the complex $x$ plane. 
In particular 
\EQ{
 \int \Omega \sim \int y\, dx\ , 
}
where $\Omega$ is integrated on a vanishing $3$-cycle in the Calabi-Yau
that reduces to one of the vanishing one-cycles on the matrix mode spectral
curve.
Putting everything together, we find that 
\EQ{
F_g \sim \Delta^{2-2g} \thicksim \left( \int y\,dx \right)^{2-2g} 
\thicksim \left( \int \Omega \right)^{2-2g} 
}
which is precisely the behaviour we expect for the free energy of the 
topological B model on the Calabi-Yau \cite{BCOV}, in agreement 
with the Dijkgraaf-Vafa correspondence.

\subsection{The double-scaling limit of $F$-terms}\label{DoubleFterms}

In this section, we will review the double-scaling limit of
various $F$-terms in the low-energy effective action
derived in \cite{BD,BHM}. 
These results were used to argue that, in the case of no double points,
the supersymmetry of the low-energy theory is actually enhanced
from four to eight supercharges.
In the next section, we will then compare some of these F-terms
with their counterparts in the corresponding ${\cal N}=2$ theory.

The effective action is written in terms of chiral superfields $S_{l}$
and $w_{\alpha l}$ which are defined as gauge-invariant single-trace  
operators \cite{CDSW}
\SP{
S_{l} & = - \frac{1}{2\pi i}\oint _{A_{l}} \, dx\, \frac{1}{32\pi^{2}} 
{\rm Tr}_{N}\left[ \frac{W_{\alpha}W^{\alpha}}{x- \Phi}\right]\ ,\\ 
w_{\alpha l} & = \frac{1}{2\pi i}\oint _{A_{l}} \, dx 
\, \frac{1}{4\pi} {\rm Tr}_{N}\left[ 
\frac{W_{\alpha}}{x- \Phi}\right] \ . 
\label{gidef}
}
It will also be convenient to define component fields for each of
these superfields, 
\EQ{
S_{l}= s_{l}+\theta_{\alpha}\chi^{\alpha}_{l}+ \cdots \ ,\qquad
w_{\alpha l}=\lambda_{\alpha l} +\theta_{\beta}f^{\beta}_{\alpha
  l}+ \cdots \ .
\label{comp}
}
The component fields,  $s_{l}$ and $f_{l}$ 
are bosonic single trace operators whilst 
$\chi_{l}$ and $\lambda_{l}$ are fermionic single trace operators. 
In the large-$N$ limit, these operators should 
create bosonic and fermionic
colour-singlet single particle states respectively. 
It is instructive to consider the interaction vertices for these
fields contained in the $F$-term effective action whose general form is
given by  \cite{Dijkgraaf:2002fc,DV3,DVPW}
\begin{equation} 
{\cal L}_{F}=
{\rm Im}\left[\int d^{2}\theta \,\big( W_\text{gb} +
  W_\text{eff}^{(2)}\big)\right]\ ,
\label{fterm0}
\end{equation}
where
\EQ{
W_\text{eff}^{(2)}  = 
\frac{1}{2}\sum_{k,l} 
\frac{\partial^{2}{F_0}}{ \partial S_{k} \partial S_{l}}
\,w_{\alpha k}w^{\alpha}_{l} \ .
\label{fterm1}
}
Expanding (\ref{fterm0}) in components on-shell, we find terms like  
\begin{equation} 
\int\, d^{2}\theta\, W_\text{eff}^{(2)} \supset 
V^{(2)}_{ij}f^{i}_{\alpha\beta}f^{\alpha\beta j}  + 
V^{(3)}_{ijk} \chi^{i}_{\alpha}f^{\alpha\beta j}\lambda^{k}_{\beta}
+V^{(4)}_{ijkl}\chi^{i}_{\alpha}\chi^{\alpha j}
\lambda^{k}_{\beta}\lambda^{\beta l}\ ,
\label{vijk}
\end{equation}
where
\begin{equation}
V^{(L)}_{i_{1}i_{2}\ldots i_{L}} =
\frac{\partial^{L} {F_0}}{\partial S_{i_{1}} \partial S_{i_{2}}
\ldots\partial S_{i_{L}}}
\label{vp}
\end{equation}
for $L=2,3,4$. 
In the large-$N$ limit, $V^{(L)}$ scales like $N^{2-L}$.
We will also consider the $2$-point vertex coming from the
glueball superpotential
\begin{equation}
\int\, d^{2}\theta\, W_\text{gb} \supset 
H^{(2)}_{ij}\chi_{\alpha}^{i}\chi^{\alpha j}\ ,
\end{equation}
where 
\begin{equation}
H^{(2)}_{ij}= 
\frac{\partial^{2} W_\text{gb} }{\partial S_{i} \partial S_{j}}\ . 
\end{equation}
The matrix $H^{(2)}_{ij}$ therefore 
effectively determines the masses of the chiral multiplets $S_{l}$.   
Note that, in the large-$N$ limit, $H^{(2)}$ scales like $N^0$.

We begin by considering the couplings $V^{(2)}_{ij}$ of the low-energy $U(1)^s$ gauge
group. Each of the $U(1)$'s is associated to one of the glueball fields
$S_i$, or equivalently the set of 1-cycles $\{A_i\}$ on $\Sigma$. 
If we ignore the $U(1)$ associated to the overall 't Hooft coupling
$S$, or the cycle $A_\infty=\sum_{i=1}^sA_i$ which can be pulled off to
infinity, the couplings of the remaining ones are simply the elements of the
period matrix of $\Sigma$. 

In order to take the $a\to0$ limit, it is
useful to choose a new basis of 1-cycles  $\{\tilde A_i,\tilde B_i\}$,
$i=1,\ldots,s-1$, which is specifically adapted to the
factorization $\Sigma\to\Sigma_-\cup\Sigma_+$. 
The subset of cycles with $i=1,\ldots,[n/2]$  
vanish at the critical point while the cycles $i=[n/2]+1,\ldots,s-1$ 
are the remaining cycles which have zero intersection with
all the vanishing cycles.  
If we define the periods on $\Sigma$
\EQ{
M_{ij} = \oint_{\tilde B_j} 
\frac{x^{i-1}}{\sqrt{\sigma(x)}}\, dx 
\ ,
\qquad
N_{ij} =
\oint_{\tilde A_j} \frac{x^{i-1}}{\sqrt{\sigma(x)}}\, dx \ ,
\label{MijNij}
}
then the period matrix, in this basis, is simply
\EQ{
\Pi=N^{-1}M\ .
}
In Appendix \ref{AppA}, we calculate the $a\to0$ limit of these matrices.
The results are summarized in \eqref{lim} and \eqref{NinvLimit}. 
Using these results, we have
\begin{equation}
\Pi\longrightarrow
\left(
\begin{array}{cc}
N^{-1}_{--} M_{--} & \quad N_{--}^{-1} M_{-+}^{(0)}
+ {\cal N} M_{++}^{(0)}
\\
0 & \quad \big( N^{(0)}_{++} \big)^{-1} M_{++}^{(0)} \\
\end{array} 
\right) \ .
\label{Pi1}\end{equation}
Let us look more closely at the structure of each
block in the above matrix. First of all, by (\ref{N--}) 
\begin{equation}
\big( N_{--}\big)_{ij} 
\thicksim 
a^{ n/2 - j }
\, f^{(N)}_{ij}( \tilde b_l) \ ,\qquad \big(M_{--}\big)_{ij} 
\thicksim 
a^{ n/2 - j }
\, f_{ij}^{(M)}( \tilde b_l) \ ,
\label{Ninv--}\end{equation}
which implies  
\begin{equation}
\big( N_{--} \big)^{-1}_{ij} \, 
\big( M_{--} \big)_{jk}
= 
f^{(N)-1}_{ij} ( \tilde b_l ) 
f^{(M)}_{jk}( \tilde b_l) 
= \Pi^{-}_{ik} ( \tilde b_l)
\ .
\label{Piminus}\end{equation}
Furthermore, since $N_{--}^{-1}$ vanishes 
in the limit $a \to 0$, we find that
\begin{equation}
N_{--}^{-1} M_{-+}^{(0)}
+ {\cal N} M_{++}^{(0)} =
N_{--}^{-1} \,M_{-+}^{(0)}
- N_{--}^{-1} \,N_{-+}^{(0)} 
\big( N_{++}^{(0)} \big)^{-1} 
M_{++}^{(0)} \rightarrow 0 \ .
\end{equation}
Therefore, the period matrix has the following block-diagonal form
in the double-scaling limit
\begin{equation}
\Pi \longrightarrow 
\left(
\begin{array}{cc}
\Pi^- & 0
\\
0 & \Pi^+ \\
\end{array} 
\right) \ .
\label{Pifinal}\end{equation} 
The upper block $\Pi^-$ is actually
the period matrix of the near-critical spectral curve $\Sigma_-$ 
\eqref{ncc} since the cycles 
$\{\tilde A_i,\tilde B_i\}$, for $i\leq[n/2]$ 
form a standard homology basis for $\Sigma_-$. Similarly, the
lower block $\Pi^+$ is the period matrix of $\Sigma_+$. So in
the limit $a\to0$ the curve $\Sigma$ factorizes as
$\Sigma_-\cup\Sigma_+$. The fact that the
period matrix factorizes is evidence of the more stringent claim that
the whole theory consists of two decoupled sectors ${\cal
  H}_-$ and ${\cal H}_+$ in the double-scaling limit.
Note that although we did not consider it, 
the $U(1)$ associated to $S$ only couples to the ${\cal H}_+$ sector.

We can extend this discussion to include other $F$-terms that are
derived from the glueball superpotential. 
For example,  consider the 3-point vertex
\EQ{
V^{(3)}_{ijk} = 
\frac{\partial^3 F_0}{\partial\tilde S_i \partial\tilde S_j
  \partial\tilde S_k}\ .
}
Here, the $\tilde S_i$ as defined as in
\eqref{defs} but with respect to the cycles $\tilde A_i$. They are
related to the $S_i$ by an electro-magnetic duality transformation.
There is a closed expression for these couplings 
of the form \cite{CMMV,Krichever,DVWDVV,Itoyama}
\begin{equation}
V^{(3)}_{ijk} 
= \frac1{N\varepsilon}\sum_{l=1}^{2s}\text{Res}_{b_l} 
\ \frac{\omega_i \,\omega_j \,\omega_k}{dx dy} \  ,
\label{d3FdS3}\end{equation}
where $\{\omega_j\}$ are the holomorphic 1-forms normalized with
respect to the basis $\{\tilde A_i,\tilde B_i\}$.
So we can deduce the behaviour of the couplings
from our knowledge of the scaling of $\omega_j$. This is derived in
Appendix \ref{AppA}. We find that the couplings are regular as $a\to0$,
except if $i,j,k\leq[n/2]$ in which case,
\EQ{
V^{(3)}_{ijk} \longrightarrow
\big(N \varepsilon a^{m+n/2+1} \big)^{-1} 
\sum_{l=1}^{n} \text{Res}_{\tilde{\sigma}_l} 
\ \frac{\tilde{\omega}_i \,\tilde{\omega}_j 
\,\tilde{\omega}_k}{d\tilde{x} dy_-} \ ,
\label{ints}
}
where the $\{\tilde{\omega}_i\}$ are the one-forms on $\Sigma_-$.
Therefore, in the double-scaling limit proposed in \eqref{limitI1}, we
find that these interactions remain finite $\sim\Delta^{-1}$, while the
other 3-point vertices $\to0$. This is yet further  
evidence of the decoupling of the Hilbert space into
two decoupled sectors where the
interactions in the ${\cal H}_-$ sector remain finite in the
double-scaling limit while those in ${\cal H}_+$ go to zero.
Notice, also that these interactions of the ${\cal H}_-$ sector 
depend universally on $\Sigma_-$.

The final $F$-term quantity that we consider is the 
Hessian matrix for the glueball superfields 
\EQ{
H^{(2)}_{jk} = 
\frac{\partial^2 W_\text{gb}}{\partial\tilde S_j 
\partial\tilde S_k} \ .
}
Using (\ref{d3FdS3}) we find 
\EQ{
H^{(2)}_{jk} = \sum_{i=1}^s N_i 
\frac{ \partial^3 F_0 }{\partial \tilde S_i\partial\tilde S_j 
\partial\tilde S_k}  
=\frac1{N\varepsilon} \sum_{l=1}^{2s} \text{Res}_{b_l} \ 
\frac{T\,\omega_j \,\omega_k}{dx dy} \ ,
\label{hess}
}
where we have defined the 1-form $T$ 
\EQ{
T=N\varepsilon\sum_{i=1}^s\frac{\partial y\,dx}{\partial S_i}\ .
}
It is known that $T$ can be can be written simply in terms of the
on-shell Seiberg-Witten curve \cite{CSWII}:
\EQ{
T= d \log(P_N+y_\text{SW})\ .
\label{Tdxexplicit}
}
In the limit $a\to0$, we can take the near-critical expressions for 
$y_\text{SW}$ in \eqref{swnc} and for $P_N(x)=2\Lambda^N$ to get the
behaviour  
\EQ{
T\longrightarrow\Lambda^{-r-n/2}H_r(0)\frac
N{n+2r}a^{n/2}d\sqrt{\tilde B(\tilde x)}\thicksim Na^{n/2}\ .
}
We also need
\EQ{
dy\longrightarrow a^{m+n/2}d\big(\tilde 
Z_m(x)\sqrt{\tilde B(\tilde x)}\big)\thicksim a^{m+n/2}\ .
} 
The scaling of the holomorphic differentials is determined in 
Appendix \ref{AppA}.

Counting the powers of $N$ and $a$, we find that for any $j$ and
$k$, $H^{(2)}_{jk}$ goes like an inverse power of $a$ and hence
diverges in the double-scaling limit (the powers of $N$ cancel). This,
however, presents us with a puzzle. In the case without double points
described in \cite{BD}, the Hessian was shown to vanish for the
${\cal H}_-$ sector, {\it i.e.\/}~$j,k\leq[n/2]$. Let us see how this is
compatible with the scaling we have just seen.
In the case, $j,k \leq [n/2]$,
\EQ{
H^{(2)}_{jk} \thicksim a^{-(m+1)} 
\sum_{l=1}^n \text{Res}_{\tilde{b}_l} \ \Big[
\frac{ d\sqrt{\tilde{B}_{n}(\tilde{x})}
\ \tilde{\omega}_j \,\tilde{\omega}_k}
{d\tilde{x} \ d \Big( \tilde{Z}_m(\tilde{x}) \,
    \sqrt{\tilde{B}_n(\tilde{x})} \Big)} 
\Big]\  ,
\label{H2--}
}
where 
\EQ{
 \tilde{\omega}_j = \frac{ \tilde{L}_j(\tilde{x})}{ \sqrt{ \tilde{B}_n(\tilde{x}) }}
 d \tilde{x} .
} 
and $\tilde{L}_j(\tilde{x})$ is a  polynomial of degree $[n/2]-1$.
Note that the differential $\tilde{\omega}_j \,\tilde{\omega}_k/d\tilde{x}$
has simple poles at $\tilde{x}=\tilde{b}_l$ on the curve $\Sigma_-$:
\EQ{
\frac{\tilde{\omega}_j \,\tilde{\omega}_k}{d\tilde{x}} = 
\frac{\tilde{L}_j(\tilde{x})
  \,\tilde{L}_k(\tilde{x})}{\tilde{B}_n(\tilde{x})}d\tilde{x} \ , 
}
but has no pole at $\tilde{x}=\infty$. For example for $n$ odd, we find  
\EQ{
\frac{\tilde{\omega}_j \,\tilde{\omega}_k}{d\tilde{x}} 
\longrightarrow \frac{d \tilde{x}}{\tilde{x}^3} \ .
}
This means that in the case with no double points, $m=0$, the 
Hessian matrix elements (\ref{H2--}) vanish identically:
\EQ{
H^{(2)}_{jk} \thicksim a^{-1} 
\sum_{l=1}^n \text{Res}_{\tilde{b}_l} \ \Big[
\frac{ \tilde{\omega}_j \,\tilde{\omega}_k}
{d\tilde{x}} \Big]
= 0 \ ,
}
because the sum of all residues of a meromorphic differential
on the compact near-critical curve $\Sigma_-$ is identically zero. 
This is precisely the result found in \cite{BD}. 
On the other hand, if $m > 0$, the Hessian matrix element 
will not vanish in general, because the differential 
on the right-hand side of (\ref{H2--}) has extra simple poles 
at the roots of
\EQ{
2 \tilde{Z}'_m(\tilde{x}) \tilde{B}_n(\tilde{x}) 
+ \tilde{Z}_m(\tilde{x}) \tilde{B}'_n(\tilde{x}) = 0 \ .
}

This result is very significant because it highlights an important
difference between the case with and without double points.
Even though we do not have control over the kinetic terms of the
glueball states, we take this behaviour of the Hessian matrix
to signal that, with double points, 
the masses of the glueball fields 
become very large in the double-scaling limit.
This is to be contrasted with the case without double points 
studied in \cite{BD},
where the appearance of the $[n/2]$ massless glueballs was interpreted as
evidence that supersymmetry is enhanced to ${\cal N}=2$
in the double-scaling limit.

\section{Matrix Model Loop Correlators}\label{loops}

In \cite{BHM}, it was shown that the large $N$ double-scaling limits defined
in \cite{BD,BHM} map to double-scaling limits of the auxiliary
Dijkgraaf-Vafa matrix model, which are completely analogous to the
``old" matrix model double-scaling limits \cite{oldMMdsl,Matrixreviews}. 
The natural question which arises is what $c \leq 1$ non-critical bosonic
string is dual to these matrix model DSLs \cite{BertoldiNC}?

In general, according to the Dijkgraaf-Vafa correspondence \cite{Dijkgraaf:2002fc,DV3,DVPW}, 
the matrix model with polynomial superpotential $W_q(\Phi)$ 
is dual to the topological B model on a non-compact Calabi-Yau geometry
which is related to the matrix model spectral curve in a simple way
\EQ{
y^2 = W'_q(x)^2 + f_{q-1}(x) \quad  \rightarrow \quad uv + y^2 =
W'_q(x)^2 + f_{q-1}(x)\ .
}
The effect of the DSL is to focus in a neighbourhood of a certain 
singularity of the above family of Calabi-Yau's parametrized by the 
superpotential couplings and deformation polynomial $f_{q-1}$. 
For instance, in the cases considered in \cite{BD} where $n$ branch points
of the matrix model spectral curve collide, we are in the proximity of a
singularity of type $A_{n-1}$ 
\EQ{
uv + y^2 = x^n - \mu\ , 
\label{singA}}
which is a generalization of the conifold singularity.
These non-compact Calabi-Yaus can generically be embedded in weighted projective
spaces, for instance \eqref{singA} goes to
\EQ{
uv + y^2 = x^n - \frac{\mu}{z^k} \ , \quad k = \frac{2n}{n+2} 
\label{eqproj}}
and have been argued to admit a Landau-Ginzburg description
with a superpotential determined by the defining equation \eqref{eqproj}
\cite{GV,OoguriVafa} (see also \cite{GhoshalMukhi,HananyOzPlesser}).
This is a generalization of the CY/LG correspondence in the compact case.
Furthermore, the superstring vacua corresponding to these generic 
non-compact CYs in the proximity of such singularities 
are expected to be described by non-critical superstring backgrounds
of the form \cite{GK,GKP}
\EQ{
\left\{ SL(2,R)/U(1) \ {\rm supercoset} \right\}  \times 
\left\{ {\cal N}=2 \ {\rm minimal \ model} \right\}  
\label{SL}
}
and their mirror symmetry partners \cite{HoriKapustin,Tong:2003ik}
\EQ{
\left\{ {\cal N}=2 \ {\rm Liouville} \right\}  \times 
\left\{ {\cal N}=2 \ {\rm minimal \ model} \right\}  
\label{LIOU}
}
In particular, in \cite{GK}, it was proposed that the four-dimensional 
double-scaled Little
String Theory with $8$ supercharges defined at such singularities 
has a holographic description in terms of the above 
non-critical superstring backgrounds. 
In fact, this correspondence is at the basis of the duality proposal
of \cite{BD}.

Therefore, since the topological B model in a neighbourhood of these CY singularities 
is dual to a topological twist of the above non-critical 
superstring backgrounds, we expect the matrix model DSL to correspond
precisely to the topological twist of these non-critical superstring backgrounds. 
For the singularities \eqref{singA}\eqref{eqproj}, we should consider
the A-twist of 
\EQ{
SL(2,R)_k/U(1) \times SU(2)_n/U(1)  \ , \quad k = \frac{2n}{n+2}  \ .
\label{SLkSUn}}

The relation between strings on non-compact Calabi-Yaus
and non-critical superstring brackgrounds \cite{OoguriVafa,GK} involving the $N=2$
Kazama-Suzuki $SL(2)/U(1)$ model or its mirror, $N=2$ Liouville theory
\cite{GK,HoriKapustin,Tong:2003ik}, has been studied by several authors
(see \cite{EguchiSugawara1,EguchiSugawara2,Israel} and references
therein).

In this section, we will study the matrix model in the double-scaling limit
and in particular derive exact expressions for its loop correlators
at genus zero by means of the algorithm developed in \cite{Eynard}.
>From the loop correlators, one can extract
correlation functions of operators of the dual $c \leq 1$ theory,
as was done for 2d gravity coupled to $(2,2m-1)$ minimal models
\cite{MSS,BanksDSS,Matrixreviews}.

Given the matrix integral
\EQ{
Z = \int d \hat\Phi \  e^{- {\hat N}\text{Tr}\,V(\hat\Phi)} \ , 
}
the $p$-loop correlator, or $p$-point loop function, is defined as
\SP{
W(x_1,\ldots,x_p) 
&\equiv {\hat N}^{p-2} \Big\langle \, \text{tr} \frac{1}{x_1-\hat\Phi}
\cdots \text{tr} \frac{1}{x_p - \hat\Phi} \, \Big\rangle_\text{conn}
\label{ploop}}
and it has the following genus expansion
\EQ{
W(x_1,\ldots,x_p) = \sum_{g=0}^\infty \frac{1}{\hat N^{2g}}
W^{(g)}(x_1,\ldots,x_p) \ .
\label{ploopgenus}}

The $1$-loop operator or matrix model resolvent 
is the Laplace transform of the macroscopic loop operator
\EQ{
W( \ell ) = \frac{1}{\hat N} \langle \, \text{Tr}\, \, e^{\ell \,\hat\Phi} \, \rangle
}
\EQ{
W(x) = \int_{0}^{\infty} d \ell \, e^{-x \ell} \, W(\ell)
= \frac{1}{\hat N} \langle \, \text{Tr}\, \frac{1}{x - \Phi} \,\rangle 
}
The macroscopic loop operator $W(\ell)$ corresponds to
the insertion of a loop of length $\ell$ on the two-dimensional 
discretized matrix model surface and encodes information on local operators 
in the dual non-critical string \cite{MSS,BanksDSS,Matrixreviews}.
In particular, 
\EQ{
W( \ell ) \, \thicksim \, \sum_{j \geq 0}  \ell^{x_j} \sigma_j  \ , \quad
x_j > 0\ , 
}
where the $\sigma_j$'s are operators in the $c  \leq 1$ system.
The correlation functions of the $\sigma_j$'s 
can then be extracted by shrinking the macroscopic loops,
namely by studying the $\ell \rightarrow 0$ 
limit of $\langle W(\ell_1) W(\ell_2) \rangle$, 
$\langle W(\ell_1) W(\ell_2) W(\ell_3) \rangle$, {\it etc\/}.

In \cite{Eynard}, Eynard found a solution to the matrix model loop
equations that allows to write down an expression for the
multiloop correlators \eqref{ploop}\eqref{ploopgenus}
at any given genus in terms of a special set of Feynman diagrams.    
The various quantities involved depend only on the spectral curve of the 
matrix model and in particular one needs to evaluate residues of certain
differentials at the branch points of the spectral curve.  

This algorithm and its extension to calculate 
higher genus terms of the matrix model free energy \cite{ChekhovEynard}
represent major progress in the solution
of the matrix model via loop equations 
\cite{ACKM,Akemann,AmbjornAkemann,Makeenko}. 
This is particularly important because, as reviewed in \cite{BHM},
the orthogonal polynomial approach can be applied 
to multi-cut solutions in very special cases only.
Another nice feature of the loop equation algorithm is that it 
shows directly how the information is encoded in the spectral curve.
This fact allowed us make precise statements about 
the double-scaling limits of multiloop correlators
and higher genus quantities simply
by studying the double-scaling limit of the spectral curve and its
various differentials \cite{BHM}.

We will now give the expression of the $2$ and $3$-loop correlators
at genus zero using Eynard's results and then consider their double-scaling
limit. Given the matrix model spectral curve for an $s$-cut solution in the
form \eqref{mc} 
\EQ{
y^2 = Z_m(x)^2 \sigma_{2s}(x)\ ,
}
the genus zero $2$-loop function is given by
\SP{
W(x_1,x_2) 
&= -\frac{1}{2(x_1-x_2)^2} +
\frac{\sqrt{\sigma(x_1)}}{2\sqrt{\sigma(x_2)}(x_1-x_2)^2}\\
&-\frac{\sigma'(x_1)}{4(x_1-x_2)\sqrt{\sigma(x_1)}\sqrt{\sigma(x_2)}}
+ \frac{A(x_1,x_2)}{4\sqrt{\sigma(x_1)}\sqrt{\sigma(x_2)}}\ .  
\label{2loop}
}
The symmetric polynomial $A$ is defined as 
\EQ{
A(x_1,x_2) = \sum_{i=1}^{2s} \frac{ {\cal L}_i(x_2) \sigma(x_1)
}{x_1-\sigma_i}\ , 
\label{A12}
}
where
\EQ{
{\cal L}_i(x_2) = \sum_{l=0}^{s-2} {\cal L}_{i,l}x_2^l =  -
\sum_{j=1}^{s-1} L_j(x_2) \int_{A_j} \frac{dx}{\sqrt{\sigma(x)}}
\frac{1}{(x-\sigma_i)}
\label{LLx}
}
and $s$ is the number of cuts. The order $s-2$ polynomials 
$L_j(x)$ enter the expression of the holomorphic one-forms
$\omega_j$ and are fixed by the requirement that these forms
are canonically normalized 
\EQ{
\omega_j = \frac{L_j(x) dx}{\sqrt{\sigma(x)}} \ , \qquad
\int_{A_k} \omega_j = \delta_{jk} \ , \qquad j,k=1,\ldots,s-1 \ .
}

\noindent
The genus zero $2$-loop function for coincident arguments is
\SP{
W(x_1,x_1)& = \lim_{x_2 \to x_1} W(x_1,x_2) = -\frac{\sigma''(x_1)}{8
\sigma(x_1)} + \frac{\sigma'(x_1)^2}{16 \sigma(x_1)^2} +
\frac{A(x_1,x_1)}{4 \sigma(x_1) }\\
&= \sum_{i=1}^{2s} \frac{1}{16 (x-\sigma_i)^2}
- \frac{\sigma_i''}{16 \sigma'_i(x-\sigma_i)}
+ \frac{{\cal L}_i(x)}{4(x-\sigma_i)}\ .
\label{2loopx1x1}
}
Another important object is the differential
\begin{equation}
dS_{2i-1}(x_1,x_2) = 
dS_{2i}(x_1,x_2)
= \frac{\sqrt{\sigma(x_2)}}{\sqrt{\sigma(x_1)}} \left(
\frac{1}{x_1-x_2} - \frac{L_i(x_1)}{\sqrt{\sigma(x_2)}} -
\sum_{j=1}^{s-1} C_j(x_2) L_j(x_1) \right) dx_1\ ,
\label{dSi}\end{equation} 
where $i=1,\ldots, s$ and 
\begin{equation}
C_j(x_2) = \int_{A_j} \frac{dx}{\sqrt{\sigma(x)} }
\frac{1}{(x-x_2)}\ .
\label{Cj}\end{equation}
A crucial aspect of the one-form (\ref{dSi}) is that it is analytic in 
$x_2$ in the limit $x_2 \to \sigma_{2i-1}$ or $\sigma_{2i}$ \cite{Eynard}
\begin{equation}
\lim_{x_2 \to \sigma_i} \frac{dS_i(x_1,x_2)}{\sqrt{\sigma(x_2)}} =
\frac{1}{\sqrt{\sigma(x_1)}} \left( \frac{1}{x_1-x_2} -
\sum_{j=1}^{s-1}  L_j(x_1) \int_{A_j} \frac{dx}{\sqrt{\sigma(x)}}
\frac{1}{(x-x_2)} \,\, \right) dx_1\ .
\label{dSilimit}\end{equation}
The subtlety is that in the definition of (\ref{Cj}),
the point $x_2$ is taken to be outside the loop surrounding the $j$-th cut,
whereas in (\ref{dSilimit}), $x_2$ is inside the contour.
Note also that
\begin{equation}
A(x_1,x_2) = - \sum_{i=1}^{2s} \left( \sum_{j=1}^{s-1} L_j(x_2)
C_j(\sigma_i) \right) \frac{\sigma(x_1)}{x_1 - \sigma_i}
\label{A12explicit}\end{equation} 
and in particular
\begin{equation}
A(x_1,\sigma_i) = {\cal L}_i(x_1) \sigma'(\sigma_i) \ .
\label{A12explicit2}\end{equation}

The expression for the genus zero $3$-loop correlator is 
$$
W_3(x_1,x_2,x_3) =
2 \sum_{i=1}^{2s} {\rm Res}_{\sigma_i} 
W_2(x, x_1) W_2(x, x_2) W_2(x, x_3)
\frac{( dx )^2}{dy}
$$
\EQ{
= \frac{1}{2} \sum_{i=1}^{2s} 
Z(\sigma_i)^2 \, \sigma'(\sigma_i) 
\ \chi_i^{(1)}(x_1)\chi_i^{(1)}(x_2)\chi_i^{(1)}(x_3)
\label{3loop}}
where the one-differentials $\chi^{(1)}_i$'s are defined by 
$$
\chi^{(1)}_i(x_1) = Res_{x \to \sigma_i} \left( \frac{dS_i(x_1,x)}{2 y(x)} \,
\frac{1}{(x-\sigma_i)} \right) 
$$
\EQ{
=
\frac{1}{2 Z(\sigma_i) \sqrt{\sigma(x_1)}} \left(
\frac{1}{x_1-\sigma_i}
+ {\cal L}_i(x_1)
\right) dx_1  
\label{chi1i}}
Incidentally, these expressions reproduce the results for
the $2$ and $3$-loop correlators in the one-cut solution
given in \cite{AJM}.

\subsection{The double-scaling limit}

As reviewed in section \ref{doublereview}, in the neighbourhood
of a singularity where $m$ double points and $n$ branch points of
the spectral curve come together
\EQ{
y^2 \rightarrow C Z_m(x)^2 B_n(x) 
}
where the double points $z_j$ and the branch points $b_i$ both tend to $x_0$, 
which we can take, without loss of generality, to be $x_0=0$. 
The double-scaling limit involves first taking $a\to0$
$$
x=a\tilde x\ ,\qquad z_i=a\tilde z_i\ ,\qquad b_j=a\tilde b_j
$$
while keeping tilded quantities fixed.
In the limit, we can define the near-critical curve
$\Sigma_-$:
\EQ{
\Sigma_-: \qquad y_-^2=\tilde Z_m(\tilde x)^2\tilde B_n(\tilde
x)\ .
\label{nccagain}
}

It was shown in \cite{BHM} that in the double-scaling limit 
\eqref{limitI1}
$$
a \rightarrow 0 \ , \qquad N \rightarrow \infty
\ , \qquad \Delta \equiv N a^{m+n/2+1} = \text{const} \ 
$$
The matrix model $p$-loop correlators behave as follows
\EQ{
W_p(x_1,\ldots,x_p) \, dx_1 \ldots dx_p  \ \rightarrow \
C^{1-p/2} \, \Delta^{2-p} \, \tilde{W}_p(\tilde{x}_1, \ldots, \tilde{x}_p) \, d \tilde{x}_1 \ldots d \tilde{x}_p
\label{ploopdsl}}
where the tilded quantities are the loop correlators corresponding
to the near-critical curve $\Sigma_-$. This result was derived by 
considering the limit of all the various differentials and quantities 
that enter in Eynard's algorithm. 

In the case where two branch points collide, which is equivalent
to a conifold singularity, we can set
\EQ{
y_-^2 =  \tilde{\sigma}(\tilde{x}) = \tilde{x}^2 - \tilde{b}^2 \ .
\label{conifold}}
Dropping the tildes and setting $C=1$, by \eqref{2loop}\eqref{ploopdsl}, the $2$-loop and
$3$-loop correlators become
\EQ{
W(x_1,x_2) = \frac{1}{2(x_1-x_2)^2} \left(
\frac{x_1 x_2 - b^2}{\sqrt{x_1^2 - b^2}{\sqrt{x_2^2 - b^2}}} - 1
\right)
\label{2loopc=1}
}
and
\EQ{
W_3(x_1,x_2,x_3) 
= \frac{1}{2 \, \Delta} \sum_{i=1}^{2} 
\, \sigma'(\sigma_i) 
\ \chi_i^{(1)}(x_1)\chi_i^{(1)}(x_2)\chi_i^{(1)}(x_3)
\label{3loopc=1}
}
where the one-differentials $\chi^{(1)}_i$'s are defined by 
\EQ{
\chi^{(1)}_i(x_1) = 
\frac{1}{2 \sqrt{x^2_1 - b^2}} \left(
\frac{1}{x_1-\sigma_i}
\right) dx_1  \ .
\label{chi1c=1}
}
The inverse Laplace transform of these genus zero correlators can be done
explicitly (see Appendix \ref{AppLaplace}) to find
\EQ{
\langle \, W( \ell_1 ) W( \ell_2 ) \, \rangle =  \sum_{n=1}^\infty 
n \, I_n( b \ell_1 ) \, I_n ( b \ell_2 ) 
\label{2ptc=1}}
\EQ{
\langle \, W( \ell_1 ) W( \ell_2 ) W( \ell_3) \, \rangle
= \frac{8}{b^2 \Delta} 
\sum_{p,q,r=1}^\infty p q r  \left( 1 + (-1)^{p+q+r} \right) I_p( b \ell_1 )\, I_q ( b \ell_2 )\, I_r( b \ell_3 )   
\label{3ptc=1}}
where $I_n(x)$ is the modified Bessel function. 

\vspace{0.2in}
\noindent
>From the correlation function of two macroscopic loop operators 
\eqref{2ptc=1}, we can extract the
wavefunction of local operators in the dual non-critical string 
\cite{BanksDSS,MSS,Matrixreviews}
\EQ{
\psi_n( \ell ) \ \sim \ \langle \, W(\ell) \, \sigma_n \,\rangle
\ \sim \  I_n(b \ell) \ .
\label{wavef}}
This wavefunction satisfies the differential equation 
\EQ{
\left( - \left( \ell \frac{\partial}{\partial \ell} \right)^2 
+ 4 \mu^2 \ell^2 + n^2 \right) \psi_n(\ell) = 0 \ ,    \quad b = 2 \mu \ .
\label{WdWelle}}
This equation corresponds to the Wheeler-DeWitt
equation of the non-critical string in the minisuperspace 
approximation, where only the zero mode $\phi_0$ of the Liouville field
is taken into account, $\ell \rightarrow e^{\gamma \phi_0/2}$ 
\EQ{
\left( - \left( \ell \frac{\partial}{\partial \ell} \right)^2 
+ 4 \mu^2 \ell^2  + \nu^2 \right) \psi_{\cal O}(\ell) = 0 \ ,
\label{WdW}}
where $\nu^2$ is related to the conformal dimension 
$\Delta^0({\cal O})$ of the undressed matter operator by
\EQ{
\nu^2 = \frac{8}{\gamma^2} \left[
\frac{Q^2}{8} - (1 - \Delta^0({\cal O}) ) 
\right] = \frac{4}{\gamma^2} \left( \alpha - \frac{Q}{2}  \right)^2
\label{nu2}}
and $\alpha$ is the Liouville charge associated to the dressing
operator $e^{\alpha \phi}$.

The Liouville background charge $Q$ and the exponent $\gamma$
are given by
\EQ{
Q = \frac{2}{\gamma} + \gamma \ , 
\quad
\gamma = \frac{1}{\sqrt{12}} \left(
\sqrt{25 - c_M} - \sqrt{1 - c_M}
\right) \ ,
\label{Liouville1}}
where $c_M$ is the central charge of the matter sector.
The Liouville central charge is
$$
c_L = 1 + 3 Q^2  \ , \quad c_L + c_M = 26 \ .
$$

The wavefunctions \eqref{wavef} are concentrated in the region
$\ell  \sim e^{\gamma\phi_0/2}\gg 1$ whereas they vanish in the $\ell \to 0$ limit. 
On the other hand, in \cite{Seibergbound,PolchinskiLFT}, 
it was argued that 
for a wavefunction to correspond to 
a physical operator in the dual $c \leq 1$ non-critical bosonic string
it should have support in the region $\ell \ll 1$, which corresponds
to infinitesimally small worldsheet metrics $e^{\gamma\phi_0} |dz|^2$. 
We can then conclude that the operators $\sigma_n$ do not
correspond to local physical observables because they do not
satisfy this requirement. The Liouville operator that dresses them
will not satisfy the Seiberg bound $\alpha \leq \frac{Q}{2}$.
Nevertheless, we will see that this is actually not a contradiction.
In fact, we expect the matrix model double-scaling limit to be
dual to the A-twist of the $N=2$ supersymmetric coset $SL(2)_k/U(1)$
at level $k=1$, the topological cigar.  
It was explicitly shown in \cite{MV} that this twisted theory,
the topological cigar, is
equivalent to the $c=1$ system at selfdual radius 
(see also \cite{NN,AshokMT,AshokT} for a recent analysis).
This result was later explained in \cite{GV} which showed the 
relation with the topological theory at a conifold singularity.
By the Dijkgraaf-Vafa correspondence \cite{Dijkgraaf:2002fc,DV3,DVPW,ADKMV}, 
the matrix model with near-critical spectral curve \eqref{conifold} 
indeed captures the topological theory of the conifold.

Using these results, we can then identify the operators $\sigma_n$ 
in the $c=1$ theory at selfdual radius. In the notation of \cite{WittenZwiebach}, 
we find that
\EQ{
\sigma_n  \  \rightarrow \  Y^-_{\frac{n}{2},\frac{n}{2}} 
= c \bar{c} \, e^{i \sqrt{2}n X_0/2} e^{\sqrt{2}(1 + n/2) \phi} \ ,
\quad n=1,2,\ldots
\label{sigman}} 
In fact from \eqref{Liouville1} we find
\EQ{
\gamma = \sqrt{2} \ , \quad Q = 2 \sqrt{2} \ , 
}
and from the Wheeler-DeWitt equation \eqref{WdWelle}
\EQ{
\nu^2 = n^2 \ , \quad n=1,2,\ldots
\label{EQnu}}
Observe that $\nu^2$, \eqref{nu2}, is invariant under
$\alpha \rightarrow Q - \alpha$. This corresponds to the fact that the
conformal dimension of the Liouville operator $e^{\alpha \phi}$ 
\EQ{
\Delta(\alpha) = \frac{1}{2} \alpha ( Q - \alpha )
\label{DimLiouville}}
is also invariant under the reflection $\alpha \rightarrow Q - \alpha$.
We also know from \eqref{wavef} that the wavefunction of the Liouville operator 
is not concentrated in the region $\ell << 1$ and therefore the corresponding 
$\alpha$ does not satisfy the Seiberg bound $\alpha \leq \frac{Q}{2} = \sqrt{2}$.
The solutions to \eqref{EQnu} compatible with this condition are
\EQ{
\alpha_n = \sqrt{2} + \frac{n}{\sqrt{2}} \ , \quad  n =1,2,\ldots
\label{alphan}}
to be contrasted with the dual solutions
$$
\tilde{\alpha}_n = Q - \alpha_n = \sqrt{2} - \frac{n}{\sqrt{2}} \ , \quad  n =1,2,\ldots
$$

The operators \eqref{sigman} indeed correspond to a subset
of the full observables in the topological cigar which
is given by
$$
Y^+_{\frac{n}{2}, -\frac{n}{2}} 
= c \bar{c} \, e^{-i \sqrt{2}nX_0/2} e^{\sqrt{2}(1 - n/2) \phi} \ ,
\quad n=0,1,2,\ldots
$$
\EQ{
Y^-_{\frac{n}{2},\frac{n}{2}} 
= c \bar{c} \, e^{i \sqrt{2} nX_0/2} e^{\sqrt{2}(1 + n/2) \phi} \ ,
\quad n=0,1,2,\ldots
\label{fullcigar}}
and their duals \cite{MV,NN,AshokMT}
$$
Y^+_{\frac{n}{2}, \frac{n}{2}} 
= c \bar{c} \, e^{i \sqrt{2}nX_0/2} e^{\sqrt{2}(1 - n/2) \phi} \ ,
\quad n=0,1,2,\ldots
$$
\EQ{
Y^-_{\frac{n}{2},-\frac{n}{2}} 
= c \bar{c} \, e^{- i \sqrt{2} n X_0/2} e^{\sqrt{2}(1 + n/2) \phi} \ ,
\quad n=0,1,2,\ldots
\label{fullcigardual}}
and also contains operators that satisfy the Seiberg bound.

\section{Comparison with ${\cal N}=2$ Seiberg-Witten Theory}\label{N=2}

In this section we will further study and discuss the enhancement to ${\cal N}=2$
supersymmetry of the ${\cal N}=1$ effective action for the case with no double points.
In particular, we will compare the double-scaling limit \eqref{limitI1} of 
the third derivatives of the matrix model free energy
\EQ{
\frac{\partial^3 F_0}{\partial S_i \partial S_j \partial S_k} 
\label{d3FlimitN=2}}
with
\EQ{
\frac{\partial^3 {\cal F}}{\partial a_i \partial a_j \partial a_k} \ ,
}
where ${\cal F}$ is the prepotential of an
${\cal N}=2$ pure $SU(n)$ Seiberg-Witten theory in the neighbourhood
of an $A_{n-1}$ Argyres-Douglas superconformal fixed point. 
It is understood that $S_i,a_i$, $i=1,\ldots,[(n-1)/2]$, are the periods of the
matrix model and Seiberg-Witten differentials around the cycles in the critical region.
This means that we have chosen the same basis of one-cycles on the
spectral and Seiberg-Witten curves as described in section \ref{DoubleFterms}. 

The goal is to provide a further consistency check that in the double-scaling limit \eqref{limitI1}
the F-terms of the large $N$ theory are equivalent to those of an ${\cal N}=2$ Seiberg-Witten 
model in the neighbourhood of an Argyres-Douglas singularity.
We will exploit exact identities that relate the third derivatives of the genus zero matrix model 
free energy and the Seiberg-Witten prepotential 
to a sum of residues on the spectral curve and Seiberg-Witten curve
\cite{DVWDVV,Marshakov:1996ae}
\AL{
\frac{\partial^3 F_0}{\partial S_i \partial S_j \partial S_k} &
= 
\sum_{l=1}^{2n}  {\rm Res}_{\sigma_l}  
\left(  \frac{\omega_i \omega_j \omega_k}{dx dy}  \right) \\
\frac{\partial^3 {\cal F}}{\partial a_i \partial a_j \partial a_k}& 
=
\sum_{l=1}^{2n}  {\rm Res}_{s_l}  
\left(  \frac{\hat\omega_i \hat\omega_j \hat\omega_k}{dx \, T_{SW}}  \right)\\ 
y^2_{SW} = P_n(x)^2 - 4 \Lambda^{2n} \ ,&
\quad
T_{SW} \equiv \frac{P'_ndx}{y_{SW}} = d \log ( P_n + y_{SW} ) 
}
As before, the $\omega_i , \hat\omega_j$'s are canonically normalized 
holomorphic one-differentials on the matrix model and 
Seiberg-Witten curves 
\EQ{
\frac{\partial ydx}{\partial S_i} = \omega_i \ ,
\qquad
\frac{\partial \lambda_{SW}}{\partial a_j} = \hat\omega_j  \  .
}
and $\sigma_l, s_l$ are the zeroes of these curves.
These formulae and their generalizations 
were used in \cite{Marshakov:1996ae,DVWDVV} to derive a set
of WDVV-like equations in Seiberg-Witten and Dijkgraaf-Vafa theories.

For the particular matrix model singularities we are interested in,
where $n$ branch points collide and there are no double-points,
the relevant matrix model spectral curve \eqref{ndp1}
\EQ{
y^2 = 
\frac1{(N\varepsilon)^2}\big(W'(x)^2+f_{\ell-1}(x)\big)
= P_n(x)^2 - 4 \Lambda^{2n} 
\label{curvan=1}}
coincides with the Seiberg-Witten curve of an $SU(n)$ theory 
\EQ{
y^2_{SW} = P_n(x)^2 - 4 \Lambda^{2n} 
\label{curvan=2}}
where
\EQ{
P_n(x) \rightarrow \frac{1}{N \varepsilon} W'_n(x) 
}
The crucial fact is that, 
in the double-scaling limit \eqref{limitI1}
\EQ{
T_{SW} \rightarrow a^{n/2} d y_-  \ , \qquad 
d y \rightarrow a^{n/2} dy_- \ ,
}
so that
$$
\frac{\partial^3 F_0}{\partial S_i \partial S_j \partial S_k} 
\rightarrow 
\left( N \varepsilon \, a^{n/2+1}  \right)^{-1} 
\sum_{l=1}^n  
{\rm Res}_{\tilde{\sigma}_l} 
\left(  \frac{\tilde\omega_i \tilde\omega_j \tilde\omega_k}{d\tilde{x} dy_-}  \right) \ ,
$$
\EQ{
\frac{\partial^3 {\cal F}}{\partial a_i \partial a_j \partial a_k} 
\rightarrow 
\left( a^{n/2+1} \right)^{-1} 
\sum_{l=1}^n  
{\rm Res}_{\tilde{\sigma}_l}  \left(  \frac{\tilde\omega_i \tilde\omega_j \tilde\omega_k}{d\tilde{x} dy_-}  \right) \  ,
\label{d3Fcompare}}
where the $\tilde\omega_i$'s, $i=1,\ldots,[(n-1)/2]$ are holomorphic 
differentials on the near-critical spectral curve
\EQ{
\Sigma_- \,: \quad y_-^2 = \tilde{B}_n(\tilde{x}) \ ,
\label{ncboh}}
and the $\tilde{\sigma}_l$'s are the $n$ zeroes of the polynomial $\tilde{B}_n(\tilde{x})$. 
We see that the third derivatives in \eqref{d3Fcompare} have exactly the
same dependence on the near-critical spectral curve.

This relation between the double-scaling limit
of a Dijkgraaf-Vafa matrix model and relative gauge theory
defined at a singularity where $n$ branch points collide 
and an ${\cal N}=2$ Seiberg-Witten theory with gauge group $SU(n)$ 
in the proximity of the analogous $A_{n-1}$ Argyres-Douglas singularity, 
which was shown to hold at genus zero, 
should extend to the higher genus F-terms as well.
In particular, if the Dijkgraaf-Vafa correspondence holds beyond the planar limit,
the double-scaling limit of the higher 
genus terms of the matrix model free energy should be related
to the higher genus Seiberg-Witten prepotentials 
of an $SU(n)$ theory close to an $A_{n-1}$ 
Argyres-Douglas superconformal fixed point. 
   
This correspondence also makes contact with the work of Nekrasov \cite{Nekrasov}, 
where it was conjectured that the full Seiberg-Witten partition 
function is actually the tau-function of a KP hierarchy and
that it is related to the theory of a chiral boson living on the Seiberg-Witten curve. 
In general, a matrix model partition function is the partition function
of a chiral boson living on the matrix model spectral curve itself and is a
tau-function of the KP hierarchy \cite{Kostov:1999xi,ADKMV}.

It is shown in \cite{BHM} that, in the double-scaling limit,
the higher genus terms of the matrix model free energy with
spectral curve $\Sigma$ behave as follows 
\EQ{
F_g ( \Sigma )  \quad \rightarrow \quad
\left(  N \varepsilon \, a^{n/2+1} \right)^{2-2g}  \ F_g ( \Sigma_- ) \ ,
\label{FgMM}} 
where $\Sigma_-$ is the near-critical spectral curve \eqref{ncboh} 
and $F_g( \Sigma_-)$ is the related genus $g$ matrix model free energy
which can be evaluated by means of the algorithms developed in 
\cite{Eynard,ChekhovEynard}.
Then the correspondence between the double-scaled matrix model 
and the ${\cal N}=2$ Seiberg-Witten theory in the neighbourhood
of an Argyres-Douglas singularity would imply that
\EQ{
{\cal F}_g ( \Sigma ) \quad \rightarrow \quad \left( a^{n/2+1} \right)^{2-2g} F_g ( \Sigma_- ) \ .
\label{FgSW}}
where $F_g( \Sigma_-)$ is again the genus $g$ matrix model free energy associated
to the near-critical spectral curve as in \eqref{FgMM}.

Based on these arguments, the Seiberg-Witten partition function in the proximity of an Argyres-Douglas
singularity would be related to the theory of a chiral boson living on the near-critical spectral curve
$\Sigma_-$.

\section{Discussion}

The analysis performed in this paper is a preliminary step towards
showing that the $c \leq 1$ non-critical bosonic string which is dual to
the matrix model double-scaling limits introduced in \cite{BD} indeed
corresponds to the topological twist of the non-critical superstring backgrounds
\eqref{SLkSUn}, which are dual to double-scaled Little String Theories
in four dimensions \cite{GK}
and to the ${\cal N}=1$ $SU(N)$ gauge theories in a partially confining phase 
in the large $N$ double-scaling limit \cite{BD}.
Using the solution of the loop equations provided by Eynard 
for a general matrix model multicut solution \cite{Eynard}, we have found
the expression of $2$ and $3$-loop matrix correlators in the DSL. 
 
More work would be needed to establish the correspondence in general.
Nevertheless, in the simplest case, where the DSL is associated to a 
conifold singularity, we have shown explicitly that the spectrum 
and wavefunctions of the operators of the non-critical bosonic string that 
can be extracted from the matrix model macroscopic loop correlators 
match with the A-twist of the $SL(2)/U(1)$ supercoset
at level $k=1$ and the $c=1$ non-critical bosonic string at selfdual radius
as expected \cite{MV,GV,OoguriVafa}.

An outstanding problem is to determine the ground ring of 
the twisted non-critical superstring background \eqref{SLkSUn} and see 
the geometry \eqref{singA} emerge from the ring relations as was done in the $c=1$ case.
The results of \cite{AshokT,EguchiSugawara1} would be particularly useful
in this respect. 
One could then couple the analysis of the ground ring 
with the study of the topological branes of \eqref{SLkSUn}
and essentially derive the matrix model
dual, as was done in \cite{SeibergShih1,SeibergShih2,SeibergShih3} 
for minimal string theories. 
Finally, it would also be interesting to study the relation between 
the matrix model DSL and a topological Landau-Ginzburg model generalizing the
analysis carried out in \cite{GhoshalMukhi,HananyOzPlesser} for the conifold/$c=1$ case.

We have also carried out another check that the F-terms of the large $N$ double-scaled
theories considered in \cite{BD} are equivalent to those of an ${\cal N}=2$ Seiberg-Witten
model in the neighbourhood of an Argyres-Douglas singularity.
This also suggests that the all-genus Seiberg-Witten partition function in a neighbourhood
of such singularities is equivalent to the double-scaled matrix model partition function
corresponding to the near-critical spectral curve \eqref{ncboh}.

\vspace{0.5in}

\noindent
{\bf Acknowledgments}. We would like to thank Nick Dorey, Prem Kumar
and Asad Naqvi for discussions.

\startappendix

\Appendix{Details of the double-scaling limit}\label{AppA}

In this appendix, we consider the double-scaling limit of various
quantities defined on the curve $\Sigma$ \eqref{mc}. This is most
conveniently done in the basis $\{\tilde A_i,\tilde B_i\}$ of 1-cycles
described in Section \ref{DoubleFterms}. In particular, for $i\leq[n/2]$ these are
cycles on the near-critical curve $\Sigma_-$ in the double-scaling
limit.   

The key quantities that we will need are the periods
\EQ{
M_{ij} = \oint_{\tilde B_j} 
\frac{x^{i-1}}{\sqrt{\sigma(x)}}\, dx 
\ ,
\qquad
N_{ij} =
\oint_{\tilde A_j} \frac{x^{i-1}}{\sqrt{\sigma(x)}}\, dx \ .
\label{MijNijapp}
}
First of all, let us focus on $N_{ij}$ where $j\leq[n/2]$, but $i$ arbitrary.
By a simple scaling argument, as $a\to0$,
\EQ{
N_{ij} 
=
\int_{b_{(j)}^-}^{b_{(j)}^+} \frac{x^{i-1}}{\sqrt{B(x)}}\, dx
\longrightarrow
a^{i-n/2}
\int_{\tilde b_{(j)}^-}^{\tilde b_{(j)}^+} 
\frac{\tilde x^{i-1}}{\sqrt{ \tilde B(\tilde{x}) }}\, d \tilde x  
= a^{i-n/2}
\, f_{ij}^{(N)} ( \tilde{b}_l) \ ,
\label{N--}
}
for some function $f_{ij}^{(N)}$ of the branch points of $\Sigma_-$.
Here, $b_{(j)}^\pm$ are the two branch points enclosed by the cycle
$\tilde A_j$. A similar argument shows that $M_{ij}$ scales  in the
same way:
\EQ{
M_{ij}\longrightarrow a^{i-n/2}
\, f_{ij}^{(M)} ( \tilde{b}_l) \ .
}
So both
$N_{ij}$ and $M_{ij}$, for $i,j,\leq[n/2]$, diverge in the limit 
$a \to 0$. On the contrary, by using a similar argument, 
it is not difficult to see that, for $j>[n/2]$, $N_{ij}$ and $M_{ij}$ 
are analytic as $a\to0$ since the integrals are over non-vanishing
cycles. 

In summary, in the limit $a \to 0$, the matrices
$N$ and $M$ will have the following block structure
\EQ{
N \longrightarrow \left(
\begin{array}{cc}
N_{--} & N_{-+}^{(0)} \\
0 & N^{(0)}_{++} \\
\end{array} 
\right)\ , 
\qquad
M \longrightarrow \left(
\begin{array}{cc}
M_{--} & M_{-+}^{(0)} \\
0 & M^{(0)}_{++} \\
\end{array} 
\right) \ ,
\label{lim}
}
where by $-$ or $+$ we denote indices in the 
ranges $\{1, \ldots, [n/2]\}$ and $\{[n/2]+1,\ldots,s-1\}$
respectively. In \eqref{lim}, $N_{--}$ and $M_{--}$ are divergent
while the remaining quantities are finite as $a\to0$.

We also need the inverse $L=N^{-1}$. 
In the text, we use the polynomials $L_j(x)=\sum_{k=1}^{s-1}L_{jk}x^{k-1}$,
which enter the expression of the 
holomorphic 1-forms associated to our basis of 1-cycles,
\EQ{
\oint_{\tilde A_i}\omega_j=\delta_{ij}\  .
}
These 1-forms are equal to
\EQ{
\omega_j(x) = \frac{L_j(x)}{\sqrt{\sigma(x)}}\, dx \ =  
\frac{\sum_{k=1}^{s-1} L_{jk} x^{k-1}}{\sqrt{\sigma(x)}}\, dx \ , \qquad
\oint_{A_i} \omega_j(x) = \delta_{ij}
\label{hold}
}
where $i,j=1,\ldots,s-1$. From the behaviour of $N$ in
the limit $a\to0$, we have
\begin{equation}
L = N^{-1}
\longrightarrow \left(
\begin{array}{cc}
N^{-1}_{--} & {\cal N} \\
0 & \left( N^{(0)}_{++} \right)^{-1} \\
\end{array} 
\right) \ ,
\qquad 
{\cal N} = - N_{--}^{-1} \, N_{-+}^{(0)} \, 
\left( N_{++}^{(0)} \right)^{-1} \ . 
\label{NinvLimit}\end{equation}
Since $N_{--}$ is singular we see that $L$ is block diagonal in the
limit $a\to0$. This is just an expression of the fact that the curve
factorizes $\Sigma\to\Sigma_-\cup\Sigma_+$ as $a\to0$. In this
limit, using the scaling of elements of $L_{jk}$, we
find, for $j\leq[n/2]$,
\EQ{
\omega_j\longrightarrow \frac{\sum_{k=1}^{[n/2]}(f^{(N)})^{-1}_{jk}\tilde
  x^{k-1}}{\sqrt{\tilde B(\tilde x)}}d\tilde x=\tilde\omega_j\ .
}
the holomorphic 1-forms of $\Sigma_-$. While for $j>[n/2]$, 
\EQ{
\omega_j\longrightarrow \frac{\sum_{k>[n/2]}^{s-1}(N_{++}^{(0)})^{-1}_{jk}
x^{k-n/2-1}}{\sqrt{F(x)}}dx\ ,
}
are the holomorphic 1-forms of $\Sigma_+$.

\Appendix{The macroscopic loop correlators for the topological cigar}\label{AppLaplace}

In this Appendix, we will evaluate the inverse Laplace trasform of
the double-scaled $2$ and $3$-loop correlators \eqref{2loopc=1}\eqref{3loopc=1}
and derive Eqs.\eqref{2ptc=1}\eqref{3ptc=1}. 

The inverse Laplace transfom of the $2$-loop correlator is given by
the double Bromwich integral
$$
\langle W(\ell_1) W(\ell_2) \rangle = 
\frac{1}{( 2\pi i )^2} \int \int  \frac{1}{2(x_1-x_2)^2} \left(
\frac{x_1 x_2 - b^2}{\sqrt{x_1^2 - b^2}{\sqrt{x_2^2 - b^2}}} - 1
\right) e^{\ell_1 x_1 + \ell_2 x_2} dx_1 dx_2 
$$
A generic $2$-loop genus zero correlator has no pole at $x_1=x_2$, \eqref{2loopx1x1}, 
and the same is true for the above integrand.
Therefore, we can deform the contours of integration and we find 
$$
\langle W(\ell_1) W(\ell_2) \rangle = 
\frac{1}{( 2\pi i )^2} \int_{A} \int_{A}  \frac{1}{2(x_1-x_2)^2} \left(
\frac{x_1 x_2 - b^2}{\sqrt{x_1^2 - b^2}{\sqrt{x_2^2 - b^2}}} - 1
\right) e^{\ell_1 x_1+ \ell_2 x_2} dx_1 dx_2
$$
where $A$ is the loop that surrounds the cut $[-b,b]$ in both
the $x_1$ and $x_2$ planes.
With the change of variables 
$$
x_i = \frac{b}{2} \left( t_i + \frac{1}{t_i} \right)
$$
$$
\langle W(\ell_1) W(\ell_2) \rangle = 
\frac{1}{( 2\pi i )^2} 
\int_{\gamma_0} \int_{\gamma_0}  \frac{1}{(1 - t_1 t_2)^2} 
e^{\frac{b}{2} \ell_1 ( t_1 + 1/t_1) + \frac{b}{2} \ell_2 ( t_2 + 1/t_2) } dt_1 dt_2
$$
where $\gamma_0$ is a counterclockwise loop around $t_i=0$.
By the identities
$$
e^{\frac{b}{2} \ell_i ( t_i + 1/t_i) }  =  \sum_{n = -\infty}^\infty I_n( b \ell_i ) \, t_i^n
$$
$$
\frac{1}{(1 - t_1 t_2 )^2} = \sum_{m=1}^\infty m ( t_1 t_2 )^{m-1}
$$
where $I_n(x)$ is the modified Bessel function, we find \eqref{2ptc=1}
\EQ{
\langle W(\ell_1) W(\ell_2) \rangle = 
\sum_{n=1}^\infty n \, I_{-n} (b \ell_1) \, I_{-n}(b \ell_2)
= \sum_{n=1}^\infty n \, I_{n} (b \ell_1) \, I_{n}(b \ell_2) \ .
}

\noindent
As for the $3$-loop correlator we can proceed in a similar manner.
First of all
$$
\chi^{(1)}_b(x_i) = \frac{1}{2 \sqrt{x^2_i - b^2}} \left(
\frac{1}{x_i - b}
\right) dx_i  = \frac{2 dt_i}{b(1 - t_i)^2} 
$$
$$
\chi^{(1)}_{-b}(x_i) = \frac{1}{2 \sqrt{x^2_i - b^2}} \left(
\frac{1}{x_i + b}
\right) dx_i  = \frac{2 dt_i}{b(1 + t_i)^2} 
$$
Since
$$
W_3(x_1,x_2,x_3) 
= \frac{1}{2 \, \Delta} \left(  
\, \sigma'(b) 
\ \chi_b^{(1)}(x_1)\chi_b^{(1)}(x_2)\chi_b^{(1)}(x_3)
+ 
\sigma'(-b) 
\ \chi_{-b}^{(1)}(x_1)\chi_{-b}^{(1)}(x_2)\chi_{-b}^{(1)}(x_3)
\right)
$$
we find 
$$
\langle \, W( \ell_1 ) W( \ell_2 ) W( \ell_3) \, \rangle
= 
$$
$$
\frac{8}{b^2 \Delta}
\frac{1}{(2\pi i)^3} \int_{\gamma_0} \int_{\gamma_0} \int_{\gamma_0}
\left( \prod_{i=1}^3 \frac{dt_i}{(1 - t_i)^2} e^{\ell_i \frac{b}{2}(t_i + 1/t_i)}
- \prod_{i=1}^3 \frac{dt_i}{(1 + t_i)^2} e^{\ell_i \frac{b}{2}(t_i + 1/t_i)}
\right) 
$$
\EQ{
= 
\frac{8}{b^2 \Delta}
 \sum_{p,q,r=1}^\infty 
p q r  \left( 1 + (-1)^{p+q+r} \right) I_p( b \ell_1 )\, I_q ( b \ell_2 )\, I_r( b \ell_3 )   \ .
\label{3ptc=1APP}}


\begin{thebibliography}{99}

{\small

\bibitem{TH1}
  G.~'t Hooft,
  ``A Planar Diagram Theory For Strong Interactions,''
  Nucl.\ Phys.\ B {\bf 72} (1974) 461.

\bibitem{AdS} J.~M.~Maldacena,
``The large-$N$ limit of superconformal field theories and supergravity,''
Adv.\ Theor.\ Math.\ Phys.\  {\bf 2} (1998) 231
[Int.\ J.\ Theor.\ Phys.\  {\bf 38} (1999) 1113]
[arXiv:hep-th/9711200].

\bibitem{BD}
  G.~Bertoldi and N.~Dorey,
  ``Non-critical superstrings from four-dimensional gauge theory,''
  JHEP {\bf 0511} (2005) 001
  [arXiv:hep-th/0507075].

\bibitem{BHM}
  G.~Bertoldi, T.~J.~Hollowood and J.~L.~Miramontes,
  ``Double scaling limits in gauge theories and matrix models,''
  JHEP {\bf 0606} (2006) 045
  [arXiv:hep-th/0603122].
    
\bibitem{Kutasov:1990ua}
  D.~Kutasov and N.~Seiberg,
  ``Noncritical Superstrings,''
  Phys.\ Lett.\ B {\bf 251} (1990) 67.    
  
\bibitem{LST}
N.~Seiberg,
``New theories in six dimensions and matrix description of M-theory on
T**5 and T**5/Z(2),''
Phys.\ Lett.\ B {\bf 408} (1997) 98
[arXiv:hep-th/9705221]. \\
M.~Berkooz, M.~Rozali and N.~Seiberg,
``On transverse fivebranes in M(atrix) theory on T**5,''
Phys.\ Lett.\ B {\bf 408} (1997) 105
[arXiv:hep-th/9704089].

\bibitem{holog}
O.~Aharony, M.~Berkooz, D.~Kutasov and N.~Seiberg,
``Linear dilatons, NS5-branes and holography,''
JHEP {\bf 9810} (1998) 004
[arXiv:hep-th/9808149].
      
\bibitem{GK}
A.~Giveon and D.~Kutasov,
``Little string theory in a double-scaling limit,''
JHEP {\bf 9910}, 034 (1999)
[arXiv:hep-th/9909110]. \\
A.~Giveon and D.~Kutasov,
``Comments on double scaled little string theory,''
JHEP {\bf 0001}, 023 (2000)
[arXiv:hep-th/9911039].

\bibitem{GKP}
A.~Giveon, D.~Kutasov and O.~Pelc,
``Holography for non-critical superstrings'',
JHEP {\bf 9910} (1999) 035
[arXiv:hep-th/9907178].
        
\bibitem{LSZLST}
  O.~Aharony, A.~Giveon and D.~Kutasov,
  ``LSZ in LST,''
  Nucl.\ Phys.\ B {\bf 691} (2004) 3
  [arXiv:hep-th/0404016].
        
\bibitem{Dijkgraaf:2002fc}
  R.~Dijkgraaf and C.~Vafa,
  ``Matrix models, topological strings, and supersymmetric gauge theories,''
  Nucl.\ Phys.\ B {\bf 644} (2002) 3
  [arXiv:hep-th/0206255].
  
\bibitem{DV3}
  R.~Dijkgraaf and C.~Vafa,
  ``On geometry and matrix models,''
  Nucl.\ Phys.\ B {\bf 644} (2002) 21
  [arXiv:hep-th/0207106].

 \bibitem{DVPW}
  R.~Dijkgraaf and C.~Vafa,
 ``A perturbative window into non-perturbative physics,''
  arXiv:hep-th/0208048.

\bibitem{oldMMdsl}
  E.~Brezin and V.~A.~Kazakov,
  ``Exactly Solvable Field Theories Of Closed Strings,''
  Phys.\ Lett.\ B {\bf 236} (1990) 144;
  M.~R.~Douglas and S.~H.~Shenker,
  ``Strings In Less Than One-Dimension,''
  Nucl.\ Phys.\ B {\bf 335} (1990) 635;
  D.~J.~Gross and A.~A.~Migdal,
  ``Nonperturbative Two-Dimensional Quantum Gravity,''
  Phys.\ Rev.\ Lett.\  {\bf 64} (1990) 127.
 
\bibitem{BertoldiNC}
  G.~Bertoldi,
  ``Double scaling limits and twisted non-critical superstrings,''
  JHEP {\bf 0607} (2006) 006
  [arXiv:hep-th/0603075].
 
\bibitem{MV}
  S.~Mukhi and C.~Vafa,
  ``Two-dimensional black hole as a topological coset model of c = 1 string
  theory,''
  Nucl.\ Phys.\ B {\bf 407} (1993) 667
  [arXiv:hep-th/9301083].

\bibitem{GV}
  D.~Ghoshal and C.~Vafa,
  ``C = 1 string as the topological theory of the conifold,''
  Nucl.\ Phys.\ B {\bf 453} (1995) 121
  [arXiv:hep-th/9506122].
  
\bibitem{OoguriVafa}
  H.~Ooguri and C.~Vafa,
  ``Two-Dimensional Black Hole and Singularities of CY Manifolds,''
  Nucl.\ Phys.\ B {\bf 463} (1996) 55
  [arXiv:hep-th/9511164].

\bibitem{ADKMV}
  M.~Aganagic, R.~Dijkgraaf, A.~Klemm, M.~Marino and C.~Vafa,
  ``Topological strings and integrable hierarchies,''
  Commun.\ Math.\ Phys.\  {\bf 261} (2006) 451
  [arXiv:hep-th/0312085].

\bibitem{BanksDSS}
  T.~Banks, M.~R.~Douglas, N.~Seiberg and S.~H.~Shenker,
   ``MICROSCOPIC AND MACROSCOPIC LOOPS IN NONPERTURBATIVE TWO-DIMENSIONAL
  GRAVITY,''
  Phys.\ Lett.\ B {\bf 238} (1990) 279.

\bibitem{MSS}
  G.~W.~Moore, N.~Seiberg and M.~Staudacher,
  ``From loops to states in 2-D quantum gravity,''
  Nucl.\ Phys.\ B {\bf 362} (1991) 665.
  
\bibitem{Matrixreviews}
  P.~Di Francesco, P.~H.~Ginsparg and J.~Zinn-Justin,
  ``2-D Gravity and random matrices,''
  Phys.\ Rept.\  {\bf 254} (1995) 1
  [arXiv:hep-th/9306153].
  P.~H.~Ginsparg and G.~W.~Moore,
  ``Lectures on 2-D gravity and 2-D string theory,''
  arXiv:hep-th/9304011.

\bibitem{Eynard}
  B.~Eynard,
  ``Topological expansion for the 1-hermitian matrix model correlation
  functions,''
  JHEP {\bf 0411} (2004) 031
  [arXiv:hep-th/0407261].

\bibitem{NN}
  S.~Nakamura and V.~Niarchos,
  ``Notes on the S-matrix of bosonic and topological non-critical strings,''
  JHEP {\bf 0510} (2005) 025
  [arXiv:hep-th/0507252].

\bibitem{AshokMT}
  S.~K.~Ashok, S.~Murthy and J.~Troost,
  ``Topological cigar and the c = 1 string: Open and closed,''
  JHEP {\bf 0602}, 013 (2006)
  [arXiv:hep-th/0511239].

\bibitem{AshokT}
  S.~K.~Ashok and J.~Troost,
  ``The topological cigar observables,''
  JHEP {\bf 0608}, 067 (2006)
  [arXiv:hep-th/0604020].

\bibitem{CSW}
  F.~Cachazo, N.~Seiberg and E.~Witten,
  ``Phases of N = 1 supersymmetric gauge theories and matrices,''
  JHEP {\bf 0302}, 042 (2003)
  [arXiv:hep-th/0301006].

\bibitem{CDSW}
  F.~Cachazo, M.~R.~Douglas, N.~Seiberg and E.~Witten,
  ``Chiral rings and anomalies in supersymmetric gauge theory,''
  JHEP {\bf 0212}, 071 (2002)
  [arXiv:hep-th/0211170].  
  
\bibitem{VY}
  G.~Veneziano and S.~Yankielowicz,
  ``An Effective Lagrangian For The Pure N=1 Supersymmetric Yang-Mills
  Theory,''
  Phys.\ Lett.\ B {\bf 113}, 231 (1982).

\bibitem{deBoer:1997ap}
  J.~de Boer and Y.~Oz,
  ``Monopole condensation and confining phase of N = 1 gauge theories via
  M-theory fivebrane,''
  Nucl.\ Phys.\ B {\bf 511} (1998) 155
  [arXiv:hep-th/9708044].
 
 \bibitem{CIV}
  F.~Cachazo, K.~A.~Intriligator and C.~Vafa,
  ``A large N duality via a geometric transition,''
  Nucl.\ Phys.\ B {\bf 603}, 3 (2001)
  [arXiv:hep-th/0103067].

\bibitem{Eguchi:2003wv}
T.~Eguchi and Y.~Sugawara,
``Branches of N = 1 vacua and Argyres-Douglas points,''
JHEP {\bf 0305} (2003) 063
[arXiv:hep-th/0305050].

\bibitem{bert}
  G.~Bertoldi,
  ``Matrix models, Argyres-Douglas singularities and double-scaling limits,''
  JHEP {\bf 0306}, 027 (2003)
  [arXiv:hep-th/0305058].

\bibitem{AD}
  P.~C.~Argyres and M.~R.~Douglas,
``New phenomena in SU(3) supersymmetric gauge theory,''
  Nucl.\ Phys.\ B {\bf 448}, 93 (1995)
  [arXiv:hep-th/9505062].

\bibitem{ARSW}
P.~C.~Argyres, M.~Ronen Plesser, N.~Seiberg and E.~Witten,
 ``New N=2 Superconformal Field Theories in Four Dimensions'',
Nucl.\ Phys.\ B {\bf 461} (1996) 71
[arXiv:hep-th/9511154].

\bibitem{EHIY}
T.~Eguchi, K.~Hori, K.~Ito and S.~K.~Yang,
``Study of $N=2$ Superconformal Field Theories in $4$ Dimensions,''
Nucl.\ Phys.\ B {\bf 471} (1996) 430
[arXiv:hep-th/9603002].

 \bibitem{Marshakov:1996ae}
  A.~Marshakov, A.~Mironov and A.~Morozov,
  ``WDVV-like equations in N = 2 SUSY Yang-Mills theory,''
  Phys.\ Lett.\ B {\bf 389} (1996) 43
  [arXiv:hep-th/9607109]; ``WDVV equations from algebra of forms,''
  Mod.\ Phys.\ Lett.\ A {\bf 12} (1997) 773
  [arXiv:hep-th/9701014];
  ``More evidence for the WDVV equations in N = 2 SUSY Yang-Mills theories,''
  Int.\ J.\ Mod.\ Phys.\ A {\bf 15} (2000) 1157
  [arXiv:hep-th/9701123].

  
\bibitem{DVWDVV}
  L.~Chekhov, A.~Marshakov, A.~Mironov and D.~Vasiliev,
  ``DV and WDVV,''
  Phys.\ Lett.\ B {\bf 562} (2003) 323
  [arXiv:hep-th/0301071].

\bibitem{CMMV}
  L.~Chekhov, A.~Marshakov, A.~Mironov and D.~Vasiliev,
  ``Complex geometry of matrix models,''
  arXiv:hep-th/0506075.

\bibitem{Krichever}
  I.~M.~Krichever,
  ``The tau function of the universal Whitham hierarchy, matrix models and
  topological field theories,''
  arXiv:hep-th/9205110.

\bibitem{Itoyama}
  H.~Itoyama and A.~Morozov,
  ``The Dijkgraaf-Vafa prepotential in the context of general Seiberg-Witten
  theory,''
  Nucl.\ Phys.\ B {\bf 657} (2003) 53
  [arXiv:hep-th/0211245].


\bibitem{BCOV}
  M.~Bershadsky, S.~Cecotti, H.~Ooguri and C.~Vafa,
  ``Kodaira-Spencer theory of gravity and exact results for quantum string
  amplitudes,''
  Commun.\ Math.\ Phys.\  {\bf 165} (1994) 311
  [arXiv:hep-th/9309140].
  
\bibitem{CSWII}
  F.~Cachazo, N.~Seiberg and E.~Witten,
  ``Chiral rings and phases of supersymmetric gauge theories,''
  JHEP {\bf 0304} (2003) 018
  [arXiv:hep-th/0303207].

 \bibitem{GhoshalMukhi}
  D.~Ghoshal and S.~Mukhi,
  ``Topological Landau-Ginzburg model of two-dimensional string theory,''
  Nucl.\ Phys.\ B {\bf 425} (1994) 173
  [arXiv:hep-th/9312189].
  
 \bibitem{HananyOzPlesser}
  A.~Hanany, Y.~Oz and M.~Ronen Plesser,
  ``Topological Landau-Ginzburg Formulation And Integrable Structure Of 2-D
  String Theory,''
  Nucl.\ Phys.\ B {\bf 425}, 150 (1994)
  [arXiv:hep-th/9401030].
  
\bibitem{HoriKapustin}
  K.~Hori and A.~Kapustin,
  ``Duality of the fermionic 2d black hole and N = 2 Liouville theory as
  mirror symmetry,''
  JHEP {\bf 0108} (2001) 045
  [arXiv:hep-th/0104202].

\bibitem{Tong:2003ik}
  D.~Tong,
  ``Mirror mirror on the wall: On two-dimensional black holes and Liouville theory,''
  JHEP {\bf 0304} (2003) 031
  [arXiv:hep-th/0303151].
 
  
\bibitem{EguchiSugawara1}
  T.~Eguchi and Y.~Sugawara,
  ``Conifold type singularities, N = 2 Liouville and SL(2,R)/U(1) theories,''
  JHEP {\bf 0501} (2005) 027
  [arXiv:hep-th/0411041].
 
\bibitem{EguchiSugawara2}
  T.~Eguchi and Y.~Sugawara,
  ``SL(2,R)/U(1) supercoset and elliptic genera of non-compact Calabi-Yau
  manifolds,''
  JHEP {\bf 0405} (2004) 014
  [arXiv:hep-th/0403193].
 
\bibitem{Israel}
  D.~Israel, A.~Pakman and J.~Troost,
  ``D-branes in N = 2 Liouville theory and its mirror,''
  Nucl.\ Phys.\ B {\bf 710} (2005) 529
  [arXiv:hep-th/0405259].

\bibitem{ChekhovEynard}
  L.~Chekhov and B.~Eynard,
  ``Hermitean matrix model free energy: Feynman graph technique for all
  genera,''
  arXiv:hep-th/0504116.

\bibitem{Makeenko}
  Y.~Makeenko,
  ``Loop equations in matrix models and in 2-D quantum gravity,''
  Mod.\ Phys.\ Lett.\ A {\bf 6} (1991) 1901.
 
\bibitem{ACKM}
  J.~Ambjorn, L.~Chekhov, C.~F.~Kristjansen and Y.~Makeenko,
  ``Matrix model calculations beyond the spherical limit,''
  Nucl.\ Phys.\ B {\bf 404} (1993) 127
  [Erratum-ibid.\ B {\bf 449} (1995) 681]
  [arXiv:hep-th/9302014].

\bibitem{Akemann}
  G.~Akemann,
  ``Higher genus correlators for the Hermitian matrix model with multiple
  cuts,''
  Nucl.\ Phys.\ B {\bf 482} (1996) 403
  [arXiv:hep-th/9606004].
 
 \bibitem{AmbjornAkemann}
  J.~Ambjorn and G.~Akemann,
  ``New universal spectral correlators,''
  J.\ Phys.\ A {\bf 29} (1996) L555
  [arXiv:cond-mat/9606129].
 
\bibitem{AJM}
  J.~Ambjorn, J.~Jurkiewicz and Yu.~M.~Makeenko,
  ``Multiloop correlators for two-dimensional quantum gravity,''
  Phys.\ Lett.\ B {\bf 251} (1990) 517. 
 
\bibitem{Seibergbound}
  N.~Seiberg,
  ``Notes on quantum Liouville theory and quantum gravity,''
  Prog.\ Theor.\ Phys.\ Suppl.\  {\bf 102} (1990) 319.

\bibitem{PolchinskiLFT}
  J.~Polchinski,
  ``Remarks on the Liouville Field Theory,'' 
  UTTG-19-90, published in Strings '90, Texas AM, College Station Wkshp (1990) 62.

\bibitem{WittenZwiebach}
  E.~Witten and B.~Zwiebach,
  ``Algebraic structures and differential geometry in $2-D$ string theory,''
  Nucl.\ Phys.\ B {\bf 377} (1992) 55
  [arXiv:hep-th/9201056].
 
 
\bibitem{Nekrasov}
  N.~A.~Nekrasov,
  ``Seiberg-Witten prepotential from instanton counting,''
  Adv.\ Theor.\ Math.\ Phys.\  {\bf 7} (2004) 831
  [arXiv:hep-th/0206161].
  
\bibitem{Kostov:1999xi}
  I.~K.~Kostov,
  ``Conformal field theory techniques in random matrix models,''
  arXiv:hep-th/9907060.

\bibitem{SeibergShih1}
  N.~Seiberg and D.~Shih,
  ``Branes, rings and matrix models in minimal (super)string theory,''
  JHEP {\bf 0402} (2004) 021
  [arXiv:hep-th/0312170].

\bibitem{SeibergShih2}
  D.~Kutasov, K.~Okuyama, J.~w.~Park, N.~Seiberg and D.~Shih,
  ``Annulus amplitudes and ZZ branes in minimal string theory,''
  JHEP {\bf 0408} (2004) 026
  [arXiv:hep-th/0406030].

\bibitem{SeibergShih3}
  N.~Seiberg and D.~Shih,
  ``Minimal string theory,''
  Comptes Rendus Physique {\bf 6} (2005) 165
  [arXiv:hep-th/0409306].


}
\end{thebibliography}
\end{document}